\documentclass[a4paper,11pt]{article}
\pdfoutput=1
\usepackage{jheppub}

\usepackage[utf8]{inputenc}

\allowdisplaybreaks

\usepackage{amssymb}
\usepackage{amsmath}
\usepackage{breqn}
\usepackage{braket}
\usepackage{xcolor}
\usepackage{graphicx}
\usepackage{slashed}
\usepackage{bbold}
\usepackage{mathrsfs}

\allowdisplaybreaks

\newcommand{\e}{\epsilon}

\newcommand{\zero}{{(0)}}
\newcommand{\one}{{(1)}}
\newcommand{\two}{{(2)}}
\newcommand{\three}{{(3)}}
\newcommand{\als}{\alpha_s}
\newcommand{\alsmu}{\alpha_s(\mu)}

\newcommand{\Ord}{\mathcal{O}}

\newcommand{\nn}{\nonumber}

\newcommand{\df}{d}

\newcommand{\Gcusp}{\Gamma^{\text{cusp}}}
\newcommand{\Lp}{L_\perp}

\def\cC{\mathcal{C}}

\def\cI{\mathcal{I}}

\DeclareMathOperator*{\SumInt}{%
\mathchoice%
  {\ooalign{$\displaystyle\sum$\cr\hidewidth$\displaystyle\int$\hidewidth\cr}}
  {\ooalign{\raisebox{.14\height}{\scalebox{.7}{$\textstyle\sum$}}\cr\hidewidth$\textstyle\int$\hidewidth\cr}}
  {\ooalign{\raisebox{.2\height}{\scalebox{.6}{$\scriptstyle\sum$}}\cr$\scriptstyle\int$\cr}}
  {\ooalign{\raisebox{.2\height}{\scalebox{.6}{$\scriptstyle\sum$}}\cr$\scriptstyle\int$\cr}}
}

\preprint{MPP-2025-161}
\title{The N$^3$LO Twist-2 Matching of TMD  Quark  Transversity}

\author[a]{Yu Jiao Zhu}
\emailAdd{yzhu@mpp.mpg.de}

\affiliation[a]{Max-Planck-Institut f\"{u}r Physik, Werner-Heisenberg-Institut, Boltzmannstr. 8, 85748 Garching, Germany}

\abstract{We present the first next-to-next-to-next-to-leading order (N$^3$LO) calculation of the twist-2 matching coefficients for  transverse momentum dependent (TMD)  quark transversity parton distribution and fragmentation functions in  QCD. This matching relates the TMD quark transversity functions to their collinear counterparts in the large-transverse-momentum regime, and provides essential ingredients for precision TMD phenomenology involving transversely polarized beams. 
As part of our analysis, we derive the next-to-next-to-leading order (NNLO)  DGLAP splitting functions for collinear transversity, confirming agreement with known space-like results for parton distribution functions and providing the new time-like splitting functions relevant for fragmentation functions.
These results extend the perturbative toolkit for spin-dependent observables and establish the transversity sector on the same theoretical footing as unpolarized and helicity distributions. Our findings enable high-precision extractions of transversity PDFs and facilitate improved theoretical predictions for azimuthal asymmetries in semi-inclusive deep inelastic scattering (SIDIS), especially in light of forthcoming data from the  Electron-Ion Collider (EIC).}

\begin{document}

\maketitle

\section{Introduction}
\label{sec:introduction}
Quark transversity distributions~\cite{Collins:2011zzd,Ralston:1979ys,Jaffe:1991kp,Jaffe:1991ra,Ji:1992ev,Jaffe:1993xb}  represent one of the three leading-twist parton distribution functions (PDFs), alongside the unpolarized and helicity distributions. 
Unlike their chiral-even counterparts, transversity distributions are chiral-odd and therefore are inaccessible through  inclusive deep inelastic scattering (DIS),
but are only  measurable  in processes involving two chiral-odd objects  that enter the cross section.
The quark transversity PDFs carry essential information about the transverse spin structure of the nucleon,
at the same time, they play a role in searches for  physics Beyond-Standard-Model (BSM) through their contribution to tensor interactions~\cite{Courtoy:2015haa}.
In particular,  precision measurements of low-energy $\beta$ decays  provide sensitive tests of possible 
BSM interactions encoded in  the effective Lagrangians describing semi-leptonic transitions with dimension-six scalar and tensor operators. 
Their contributions interfere linearly with the Standard Model amplitude, making them directly proportional to the nucleon tensor charge $g_T$~\cite{Cirigliano:2013xha},
\begin{align}
g_T = \delta^u(Q^2) - \delta^d(Q^2), \qquad
\delta^q(Q^2) = \int_0^1 \mathrm{d} x \,[h_1^q(x,Q^2)-h_1^{\bar q}(x,Q^2)] ,
\end{align}
i.e., the isovector combination of  first Mellin moments of up and down quark transversity. 
Tensor charges  are also  indispensable inputs to  nucleon electric dipole moment (EDM), 
which provide a unique window to hadronic and semi-leptonic CP violation~\cite{Yamanaka:2017mef}. 

Over the past decades, significant progress has been made in unraveling the quark transversity, 
driven by increasingly refined theoretical frameworks and experimental observations~\cite{Efremov:1992pe,Collins:1993kq,Jaffe:1997hf,Qiu:1998ia,Bianconi:1999cd,Radici:2001na,Barone:2001sp,Bacchetta:2002ux}.
Its extraction has been made possible 
either within TMD framework through  Collins effect~\cite{Collins:1992kk,Anselmino:2015sxa,Kang:2015msa,Lin:2017stx,DAlesio:2020vtw,Cammarota:2020qcw,
Gamberg:2022kdb,Boglione:2024dal},
or within collinear framework  through the mechanism of  dihadron productions~\cite{Radici:2018iag,Benel:2019mcq,Cocuzza:2023oam}. 
The first experimental evidence for transversity was obtained in semi-inclusive deep-inelastic  scattering (SIDIS)  with a transversely polarized proton target, 
$\ell+ P^{\uparrow} \to \ell’ + h + X$, by the HERMES Collaboration~\cite{HERMES:2004mhh}.
Later, both experimental analyses and theoretical studies advanced significantly,
in particular through  global analyses of SIDIS  Collins modulation  in conjunction with the chiral-odd Collins fragmentation function,
 as well as hadron-in-jet Collins eﬀect and dihadron correlations~\cite{Yuan:2007nd,Burkardt:2008jw,Anselmino:2007fs,
COMPASS:2008isr,HERMES:2010mmo,COMPASS:2010hbb,Courtoy:2012ry,Martin:2014wua,COMPASS:2014bze,STAR:2017wsi,Anselmino:2013vqa,Kang:2014zza,Cocuzza:2023vqs,COMPASS:2023vhr,COMPASS:2023vqt,Zeng:2024gun,STAR:2025xyp,LHCspin:2025lvj,Gao:2025evv,DAlesio:2010sag,DAlesio:2025jmr}. 
Recently, proposals have been put forward to probe quark transversity in SIDIS by measuring the azimuthal asymmetry of the energy flow around a tagged hadron~\cite{Cao:2025icu}. 
This approach introduces Collins-type Fragmentation Energy Correlators as an analogue of the Collins function and is formulated within collinear factorization. 
It is practically appealing because quark transverse-spin information can be extracted solely by reanalyzing existing SIDIS data, without requiring new dedicated measurements.
In parallel, transversity distributions are also independently computed in Lattice simulations~\cite{Alexandrou:2018eet,Alexandrou:2016jqi,Chen:2016utp,HadStruc:2021qdf,LatticeParton:2022xsd,Gao:2023ktu}.
 For a comprehensive review of recent progress, see Refs.~\cite{Aidala:2012mv,Boussarie:2023izj}.
 
 The  large $p_T\gg \Lambda_{\text{QCD}}$ asymptotics has provided additional theoretical constraint relating  TMD-based approach and the collinear framework.
 Significant progress has been made in recent years in developing the TMD formalism and computing perturbative ingredients at next-to-next-to-next-to-leading order (N$^3$LO) for unpolarized and helicity TMDs~\cite{Luo:2019bmw,Luo:2020epw,Ebert:2020qef,Ebert:2020yqt,Zhu:2025gts}. 
However, the precision theory of transversely polarized TMDs has lagged behind, with matching coefficients  only known up to NLO~\cite{Gutierrez-Reyes:2017glx}. The work of Gutierrez-Reyes, Scimemi, and Vladimirov~\cite{Gutierrez-Reyes:2018iod} established the NNLO matching of the transversity TMDPDF and TMDFF.
Their analysis also revealed interesting structures, such as the vanishing of the pretzelosity matching coefficients at two loops, 
suggesting that it does not match onto any twist-2 collinear distribution.
This feature is  consistent with   quark-model studies indicating that  the TMD pretzelosity distribution $h_{1T}^{\perp q}$ is directly 
related to the higher-twist quark angular momentum~\cite{She:2009jq,Avakian:2009jt,Avakian:2010br}. 
The pioneering work of Gutierrez-Reyes, Scimemi, and Vladimirov has significantly advanced the theoretical understanding of quark transversity. 
Nevertheless, we find minor discrepancies compared to their results, as discussed in detail in the main body of the manuscript.

In this work, we extend the perturbative understanding of transversely polarized TMDs to the next level of precision. 
We present the first computation of the N$^3$LO twist-2 matching coefficients for TMD quark  transversity distributions and fragmentation functions,
using an exponential regulator~\cite{Li:2016axz}.
As an outcome, we derive  the full set of NNLO  DGLAP splitting functions in analytic form for quark transversity, 
and find full agreement when cross-checking with the existing space-like results in Ref.~\cite{Blumlein:2021enk}.
Together, these developments provide a solid foundation for  global analyses of spin-dependent data and for interpreting precision measurements at the forthcoming EIC.
\section{The transversely polarized quark TMD PDFs and FFs}
\subsection{Operator definitions for   TMD quark   transversity  PDFs and FFs}
\label{sec:defin-transversity}
The transversely polarized  TMD  distribution   can be defined in terms of SCET~\cite{Bauer:2000ew,Bauer:2000yr,Bauer:2001yt,Bauer:2002nz,Beneke:2002ph} collinear fields
\begin{align}
  \label{eq:PDFdef}
   \mathcal{B}_{q/N}^{\mu}(x,b_\perp,\vec s_T) = &
\int \frac{db^-}{2\pi} \, e^{-i x b^- P^{+}} 
 \langle N(P),\vec  s_T| \bar{\chi}_n(0,b^-,b_\perp) \frac{\slashed{\bar{n}} \gamma^{\mu}_{\perp}}{2}\gamma_5 \chi_n(0) | N(P) ,\vec  s_T\rangle \,,
 \end{align}
where $\gamma^{\mu}_{\perp}$ is given by  $\gamma^{\mu}_{\perp}\equiv g^{\mu\nu}_{\perp}\gamma_{\nu}$,
 and $N(P)$ is a hadron state with momentum $P^\mu = (\bar{n} \cdot P) n^\mu/2 = P^+ n^\mu/2$, with $n^\mu = (1, 0, 0, 1)$ and $\bar{n}^\mu = (1, 0, 0, -1)$,
 $\chi_n = W_n^\dagger \xi_n$ is the gauge invariant collinear quark field~\cite{Bauer:2001ct} in SCET, 
constructed from collinear quark field $\xi_n$ and path-ordered collinear Wilson line $W_n(x) = {\cal P} \exp \left(i g \int_{-\infty}^0 ds\, \bar{n} \cdot A_n (x + \bar{n} s) \right)$.
For a polarized hadron with transverse spin $\vec{s_T}$ and mass $M$, 
the  transversely polarized TMD  distribution  is parameterized as~\cite{Goeke:2005hb,Bacchetta:2006tn,Boer:2011xd,Scimemi:2018mmi}
\begin{align}
\mathcal{B}_{q/N}^{\mu}(x,b_\perp,\vec { s_T}) =& \vec  s_T^\mu h_1(x,b_\perp)+i\e_{\perp}^{\mu\nu}b_{\perp\,\nu} M h_1^{\perp}(x,b_\perp)+\frac{M^2 b_T^2}{2}
\left(\frac{g^{\mu\nu}_{\perp}}{2}+\frac{b_{\perp}^{\mu}b_{\perp}^{\nu}}{b_T^2}\right) \vec { s_T}_\nu h_{1T}^{\perp}(x,b_\perp)\,.
\end{align}
The physical meanings of the relevant  TMDs are summarized as follows.
The  chiral-odd  Boer-Mulders function~\cite{Boer:1997nt} $h_1^{\perp}(x,b_\perp)$
 describes the distribution of transversely polarized quarks in an unpolarized target, this function is studied at NLO in~\cite{Rein:2022odl}.
It is a T-odd function just like the  Sivers function~\cite{Sivers:1989cc}, 
both are expected to change sign 
when going from SIDIS to the Drell-Yan process~\cite{Collins:2002kn,Brodsky:2002rv,Brodsky:2002cx,Brodsky:2013oya,Belitsky:2002sm}.  
Their existence is attributed to initial-state or final-state interactions between the active partons  and the spectators.
The pretzelosity distribution $ h_{1T}^{\perp}(x,b_\perp)$  has  vanishing twist-2 matching coefficients,
 as  found in~\cite{Gutierrez-Reyes:2018iod}.
The quark may contribute a transverse angular momentum proportional to the transverse spin of the nucleon target,
this contribution is described by the TMD quark  transversity $h_1(x,b_\perp)$.
In this work, we consider perturbative matching of $h_1(x,b_\perp)$ onto transversity PDFs
\begin{align}
\vec { s_T}^\mu h_1 (x) \equiv 
\int \frac{db^-}{2\pi} \, e^{-i x b^- P^{+}} 
 \langle N(P),\vec { s_T} | \bar{\chi}_n(0,b^-,0) \frac{\slashed{\bar{n}} \gamma^{\mu}_{\perp}}{2}\gamma_5 \chi_n(0) | N(P) ,\vec { s_T}\rangle \,,
\end{align}
the definition above utilizes the consequence of rotation and parity symmetries of the operator matrix elements.
The   transversity parton distribution $h_1 (x)$  quantifies the difference between the densities of quarks with transverse spin aligned and anti-aligned with the transverse spin of the parent hadron
\begin{align}
h_1 (x) =&\text{density of quark  of spin parallel to target}
 \nn\\
 -&\text{density of quark of spin antiparallel to target}\,,
\end{align}
where the target is fully transversely polarized with respect to the beam axis.

The OPE relation at small $b_T$ is
\begin{align}
h_1^q(x,b_\perp)=\int_x^1\frac{dy}{y} \sum_{f=q,\bar q} \delta \mathcal{I}_{qf}\left(\frac{x}{y},b_\perp\right) h^f_1 (x)+\mathcal O(b_T^2 \Lambda^2_{\text{QCD}})\,,
\end{align}
where the sum over flavors is restricted to non-singlet combinations, as there is no gauge-invariant operator corresponding to a transversely polarized gluon.
The matching coefficients $ \delta \mathcal{I}_{qf}$ describes physics above  the  $\Lambda_{\text{QCD}}$ scale,
as a result, it can be computed from   partonic TMD transversity PDFs    
\begin{align}
  \label{eq:PDFdef}
  \delta h_{q/j}^{\rm bare}(x,b_\perp) = &
\int \frac{db^-}{2\pi} \, e^{-i x b^- P^{+}} 
 \langle P, j_\sigma | \bar{\chi}_n(0,b^-,b_\perp) \frac{\slashed{\bar{n}} \gamma^{\mu}_{\perp}}{2}\gamma_5 \chi_n(0) | P,j_\rho  \rangle  \bar\Gamma_{\mu}^{\sigma\rho}\, ,
 \end{align}
 where the dual spin projector
 $\bar\Gamma_\mu= \frac{\gamma_5\gamma_{\perp\mu}\slashed{P} }{2}$ 
 is introduced. It is dual to
 $ \Gamma_\mu=\frac{\slashed{\bar{n}} \gamma_{\perp\mu}}{2}\gamma_5$ and is used to project out  $ \delta h_{q/j}(x,b_\perp)$, 
 the partonic version of $h_1^q(x,b_\perp)$.
 By inserting a complete set of $n$-collinear state $\mathbb{1} = \SumInt_{X_n}  \! d {\rm PS}_{X_n}      | X_n \rangle \langle X_n |$ into the operator definition,
the bare  TMD PDFs can be computed from splitting amplitudes integrated over collinear phase space  with a rapidity cutoff $\nu=1/\tau$,
implemented through the exponential regulator $e^{ -b_0 \tau \frac{P \cdot K}{P^+}}$~\cite{Li:2016axz,Luo:2019hmp} 
\begin{align}
\label{eq:amp-id1}
\delta h_{q/j}^{{\rm bare}} (x, b_\perp)= \lim_{\tau\to0}&\SumInt_{X_n}  \! d {\rm PS}_{X_n} 
e^{- i K_\perp \!\cdot b_\perp} 
e^{ -b_0 \tau \frac{P \cdot K}{P^+}}
\delta(K^+ - (1 - x) P^+) 
   \delta{\rm \bold P}^{\sigma \rho}_{\mu\, q\leftarrow j}
   \bar\Gamma^{\mu}_{\sigma \rho}\,,
\end{align}
where $K^\mu$ is the total momentum of $|X_n \rangle$, and $d\text{PS}_{X_n}$ is the collinear phase space measure,
and $\delta{\rm \bold P}^{\sigma \rho}_{\mu\,q\leftarrow j}$ is the space-like spin correlator for the target  
\begin{align}
\label{eq:sp-id}
 \delta{\rm \bold P}^{\sigma \tau}_{\mu\,q \leftarrow j}
  \equiv
  {\rm \bold{Sp}}_{X_n q_{l}^* \leftarrow j_\sigma}^{*} 
  \Gamma_{\mu}^{l s}
  {\rm \bold{Sp}}_{X_n q_{s}^* \leftarrow j_\rho} ^{}
\equiv
 \langle j_\sigma | \bar{\chi}_n| X_n \rangle   \frac{\slashed{\bar{n}} \gamma_{\perp\mu}}{2}\gamma_5  \langle X_n |  \chi_n| j_\rho\rangle\,,
\end{align}
where $  {\rm \bold{Sp}}_{X_n q_{s}^* \leftarrow j_\rho} ^{}= \langle X_n |  \chi^s_n| j_\rho\rangle$ denotes the  quark splitting amplitude with an incoming partonic target $j_\rho$ and an off-shell active quark $q_{s}^*$,
 associated with the
collinear quark source operator  $ \chi^s_n$. The indices $\rho$ and $s$ label the quark spin.
In addition, since the distribution is non-vanishing only in the non-singlet (ns) sectors, 
one may expect an equivalent definition of the integrand without the explicit appearance of $\gamma_5$. 
Our strategy is therefore to eliminate $\gamma_5$ by deforming the operator definitions in four dimensions, as follows:
\begin{align}
\frac{\slashed{\bar{n}} \gamma^{\mu}_{\perp}}{2} \gamma_5\dots \times\frac{\gamma_5\gamma_{\perp\mu}\slashed{P} }{2}\dots
=
-i \frac{1}{2} \varepsilon_{\perp}^{\mu\nu} \slashed{\bar n}\gamma_{\perp\nu}\dots\times i  \frac{1}{2} \varepsilon_{\mu\sigma} \gamma_{\perp}^{\sigma} \slashed{ P}\dots
=\frac{\slashed{\bar{n}} \gamma^{\mu}_{\perp}}{2} \dots \times\frac{\gamma_{\perp\mu}\slashed{P} }{2}\dots\,,
\end{align}  
note that it is not assumed here that  various traces are topologically connected.
With this method we have also verified that  terms proportional to the cubic color structure $d^2_{abc}$  vanish identically.
The TMD  quark transversity FFs~\cite{Metz:2016swz} can be obtained from crossing, 
and frame-dependence must be carefully taken into account~\cite{Collins:2011zzd,Luo:2019hmp,Luo:2019bmw}.
For a final state detected hadron $N$ carrying momentum $P^\mu = (\bar n \cdot P) n^\mu/2 =  P^+ n^\mu/2$ and transverse spin $\vec {s_T}$, we define
\begin{align}
  \label{eq:FF_hadron_Frame}
\delta{\cal D}_{N/q}^{\mu} (z, b_\perp/z,\vec {s_T})  =& z^{1 - 2 \e} \SumInt_{X_n}    \int \frac{db^-}{2\pi}   e^{iP^+  b^-  / z }  
  \langle 0 | 
 \bar \chi_{n}(0,b^-,b_\perp) |  N(P),\vec {s_T},X \rangle 
 \nn\\
\times&\frac{\slashed{\bar{n}} \gamma^{\mu}_{\perp}}{2}\gamma_5  \langle  N(P),\vec {s_T},X | \chi_{n}(0) | 0 \rangle  \,,
\end{align}
where again we only consider contributions proportional to the detected hadron's transverse spin, i.e., the TMD transversity FFs.
 Similar to Eq.~(\ref{eq:amp-id1}), we can define the  partonic TMD transversity FFs as 
 \begin{align}
\label{eq:amp-id2}
\delta f_{i/q}^{{\rm bare}} (z, b_\perp/z)=  z^{1 - 2 \e}\lim_{\tau\to0}&\SumInt_{X_n}  \! d {\rm PS}_{X_n} 
e^{- i K_\perp \!\cdot b_\perp} 
e^{ -b_0 \tau \frac{P \cdot K}{P^+}}
\delta\left(K^+ - \left(\frac{1}{z}-1\right) P^+\right) 
   \delta{\rm \bold P}^{T,\sigma \rho}_{\mu\,i \leftarrow q}
   \bar\Gamma^\mu_{\sigma \rho}\,,
\end{align}
 where $ \delta{\rm \bold P}^{T,\sigma \rho}_{\mu\,i \leftarrow q}$ is the square of the time-like splitting amplitude,
 which can be obtained from the space-like ones in Eq.~(\ref{eq:sp-id})  by analytical continuation.
 The  dual spin projectors $\bar \Gamma$  in Eq.~(\ref{eq:amp-id2})
 are formally identical to those of the space-like ones.

\subsection{Collinear mass factorization and renormalization group equations}
\label{sec:massFac-RG}
The bare TMD helicity PDFs (or FFs) contain both ultraviolet (UV) and infrared (IR) divergences. The procedure of UV renormalization, zero-bin subtraction, and mass factorization is technically identical to that of the unpolarized case~\cite{Luo:2020epw},
and are summarized in the following collinear mass factorization formula
\begin{align}
  \label{eq:mass-fac-form}
\frac{1}{Z^B}  \frac{\delta h_{q/j}^{{\rm bare}}(x,b_\perp)}{\mathcal{S}_{0 \rm b} } = & \sum_k \delta\mathcal{I}_{q k}(x,b_\perp,\mu,\nu) \otimes \delta\phi_{kj}(x,\mu)  \,,
\nn\\
\frac{1}{Z^B}  \frac{\delta f_{i/q}^{{\rm bare}}(z,b_\perp/z)}{\mathcal{S}_{0 \rm b} } = & \sum_k  \delta d_{ik}(z,\mu)\otimes\delta \mathcal{C}_{kq} (z, b_\perp/z,\mu,\nu)  \,,
\end{align}
where  the  symbol $\otimes$ denotes the convolution of two functions, defined as  $f(x\,,\dots)\otimes g(x\,,\dots)\equiv \int_x^1\mathrm{d} \xi/\xi \,,f(\xi\,,\dots)g(x/\xi\,,\dots)$.
 The  bare zero-bin soft function $\mathcal{S}_{0 \rm b}(\alpha_s)$   is the same as the TMD soft function~\cite{Li:2016ctv}.
 $Z^B$ (see  Sec.~\ref{sec:RC}) are the multiplicative operator renormalization constants for  TMD  PDFs and FFs , and $\delta\mathcal{I}_{qj}$ ($\delta\mathcal{C}_{iq}$) are the finite coefficient functions.
$\delta\phi_{ki}$ (and $\delta d_{ik}$) are the  partonic transversity PDFs (and FFs), 
they evolve with the transversity splitting functions $\delta P$~\cite{Vogelsang:1997ak,Mikhailov:2008my}, 
 up to $\alpha_s^3$, they are given by
 \begin{align}
\delta\phi_{ij}(x, \alpha_s) 
=& \delta_{ij} \delta(1-x) - \frac{\alpha_s}{4 \pi} \frac{{\delta P}^\zero_{ij}(x)}{\epsilon} 
\nn\\
+&  \left(\frac{\alpha_s}{4 \pi}\right)^2 \bigg[ \frac{1}{2 \epsilon^2} \biggl( \sum_k {\delta P}^\zero_{ik} \otimes {\delta P}^\zero_{k j}(x) + \beta_0 {\delta P}^\zero_{ij} (x)  \biggl) - \frac{1}{2\epsilon} {\delta P}^\one_{ij}(x)  \bigg]  
\nn \\
+&  \left(\frac{\alpha_s}{4 \pi}\right)^3 \bigg[  \frac{-1}{6 \epsilon^3} \biggl(  \sum_{m \,, k }{\delta P}^{\zero}_{im} \otimes {\delta P}^{\zero}_{mk} \otimes {\delta P}^{\zero}_{k j}(x)  + 3 \beta_0 \sum_k {\delta P}^{\zero}_{ik} \otimes {\delta P}^{\zero}_{kj}(x)
\nn \\ 
 +&2 \beta_0^2 {\delta P}^{\zero}_{ij}(x) \biggl)  
+ \frac{1}{6 \epsilon^2} \biggl( \sum_k {\delta P}^{\zero}_{ik} \otimes {\delta P}^{\one}_{kj}(x) 
+ 2 \sum_k {\delta P}^{\one}_{ik} \otimes {\delta P}^{\zero}_{kj} (x) 
\nn\\
 +& 2  \beta_0 {\delta P}^{\one}_{ij}(x) +  2  \beta_1 {\delta P}^{\zero}_{ij}(x) \biggl)
 -\frac{1}{3 \epsilon} {\delta P}^\two_{ij}(x) \bigg]
 +\Ord(\alpha_s^3)  \,,
\end{align}
 and for  the FFs
\begin{align}
\delta d_{ij}(z, \alpha_s) 
=& \delta_{ij} \delta(1-z) - \frac{\alpha_s}{4 \pi} \frac{{\delta P}^{T\zero}_{ij}(z)}{\epsilon} 
\nn\\
+&  \left(\frac{\alpha_s}{4 \pi}\right)^2 \bigg[ \frac{1}{2 \epsilon^2} \biggl( \sum_k {\delta P}^{T\zero}_{ik} \otimes {\delta P}^{T\zero}_{k j}(z) + \beta_0 {\delta P}^{T\zero}_{ij} (z)  \biggl) - \frac{1}{2\epsilon} {\delta P}^{T\one}_{ij}(z)  \bigg]  
\nn \\
+&  \left(\frac{\alpha_s}{4 \pi}\right)^3 \bigg[  \frac{-1}{6 \epsilon^3} \biggl(  \sum_{m \,, k }{\delta P}^{T\zero}_{im} \otimes {\delta P}^{T\zero}_{mk} \otimes {\delta P}^{T\zero}_{k j}(z)  + 3 \beta_0 \sum_k {\delta P}^{T\zero}_{ik} \otimes {\delta P}^{T\zero}_{kj}(z) 
\nn \\ 
+&2 \beta_0^2 {\delta P}^{T\zero}_{ij}(z) \biggl)  
+ \frac{1}{6 \epsilon^2} \biggl( 2\sum_k {\delta P}^{T\zero}_{ik} \otimes {\delta P}^{T\one}_{kj}(z) 
+  \sum_k {\delta P}^{T\one}_{ik} \otimes {\delta P}^{T\zero}_{kj} (z) 
\nn\\
 +& 2  \beta_0 {\delta P}^{T\one}_{ij}(z) 
+  2  \beta_1 {\delta P}^{T\zero}_{ij}(z) \biggl)
 -\frac{1}{3 \epsilon} {\delta P}^{T\two}_{ij}(z) \bigg]
 +\Ord(\alpha_s^3)  \,,
\end{align}
where ${\delta P}_{ij}^{(n)}$ is the $\text{N}^n$LO space-like transversity splitting function~\cite{Artru:1989zv,Barone:1997fh},
which is presently  known to NLO~\cite{Vogelsang:1997ak,Mikhailov:2008my},
and ${\delta P}_{ij}^{T (n)}$ is the $\text{N}^n$LO time-like transversity splitting function.

The factorized finite coefficient functions obey the following $\mu$-RG equations
\begin{align}
\frac{\df}{\df \ln\mu} \delta \cI_{ji}^{ }(x,b_\perp,\mu,\nu) = 2 \bigg[ \Gcusp_j(\alsmu) \ln\frac{\nu}{x P_{+}} +& \gamma^B_j(\alsmu) \bigg] \delta \cI_{ji}^{}(x,b_\perp,\mu,\nu)
\nn\\
- &2 \sum_k \delta\cI_{jk}^{}(x,b_\perp,\mu,\nu) \otimes {\delta P}_{ki}(x,\alsmu) \,,
\end{align}
\begin{align}
\frac{\df}{\df \ln\mu}\delta \cC^{ }_{ij}(z,b_\perp/z,\mu,\nu) = 2 \bigg[ \Gcusp_j(\alsmu) \ln\frac{z\nu}{ P_{+}} +& \gamma^B_j(\alsmu) \bigg] \delta\cC^{}_{ij}(z,b_\perp/z,\mu, \nu)
\nn\\
-& 2 \sum_k {\delta P}^T_{ik}(z,\alsmu) \otimes \delta\cC^{}_{kj}(z,b_\perp/z,\mu,\nu) \,,
\label{eq:Imu}
\end{align}
and the rapidity evolution equations~\cite{Chiu:2011qc,Chiu:2012ir}
\begin{align}
\frac{\df}{\df\ln\nu}\delta \cI_{ji}^{}(x,b_\perp,\mu,\nu) =& -2 \left[ \int_{\mu}^{b_0/b_T} \frac{\df\bar{\mu}}{\bar{\mu}} \Gcusp_j(\alpha_s(\bar{\mu})) + \gamma^R_j(\als(b_0/b_T)) \right] \delta\cI_{ji}^{}(x,b_\perp,\mu,\nu) \, ,
\nn\\
\frac{\df}{\df\ln\nu}\delta \cC^{}_{ij}(z,b_\perp/z,\mu,\nu) =& -2 \left[ \int_{\mu}^{b_0/b_T} \frac{\df\bar{\mu}}{\bar{\mu}} \Gcusp_j(\alpha_s(\bar{\mu})) + \gamma^R_j(\als(b_0/b_T)) \right] \delta\cC^{}_{ij}(z,b_\perp/z,\mu, \nu) \,.
\label{eq:Inu}
\end{align}
We define the coefficient function order-by-order  in terms of  strong coupling $\alpha_s(\mu)/(4 \pi)$ as
$\delta\cI_{ji}(x,b_\perp,\,\mu,\nu)=\sum_{n=0} (\alpha_s(\mu)/(4 \pi))^n\delta\cI^{(n)}_{ji}(x,b_\perp,\,\mu,\nu)$.
Truncating the perturbative expansion at $\Ord(\alpha_s^3)$, the solution to these evolution equations reads
\begin{align}
\label{eq:RGs-1}
\delta\cI^\zero_{ji}(x,b_\perp,&\,\mu,\nu) = \delta_{ji} \delta(1-x) \, ,\nn
\\
\delta\cI^\one_{ji}(x,b_\perp,&\,\mu,\nu) = \left( - \frac{\Gcusp_0}{2} \Lp L_Q + \gamma_0^B \Lp + \gamma_0^R L_Q \right) \delta_{ji} \delta(1-x) - {\delta P}_{ji}^\zero(x) \Lp +\delta I_{ji}^\one(x) \, , \nn
\\
\delta\cI^\two_{ji}(x,b_\perp,&\,\mu,\nu) =  \bigg[ \frac{1}{8} \left( -\Gcusp_0 L_Q + 2\gamma^B_0 \right) \left( -\Gcusp_0 L_Q + 2\gamma^B_0 + 2\beta_0 \right) \Lp^2
\nn\\
+& \left(  (-\Gcusp_0 L_Q + 2\gamma^B_0 + 2\beta_0) \frac{\gamma_0^R}{2} L_Q 
-\frac{\Gcusp_1}{2} L_Q + \gamma^B_1  \right) \Lp 
\nn\\
+& \frac{(\gamma_0^R)^2}{2} L_Q^2 + \gamma_1^R L_Q \bigg] \, \delta_{ji} \delta(1-x) 
+ \bigg( \frac{1}{2} \sum_l {\delta P}^\zero_{jl} \otimes {\delta P}^\zero_{li}(x) \nn
\\
+& \frac{{\delta P}^\zero_{ji}(x)}{2} (\Gcusp_0 L_Q - 2\gamma_0^B - \beta_0) \bigg) \Lp^2 + \bigg[ -{\delta P}^\one_{ji}(x) - {\delta P}^\zero_{ji}(x) \gamma_0^R L_Q 
\nn\\
- &\sum_l  \delta I^\one_{jl} \otimes {\delta P}^\zero_{li}(x) \nn
+ \left( -\frac{\Gcusp_0}{2} L_Q + \gamma_0^B + \beta_0 \right) \delta I^\one_{ji}(x) \bigg] \Lp 
\nn\\
+& \gamma_0^R L_Q \delta I^\one_{ji}(x) + \delta I^\two_{ji}(x) \, ,\nn
\\
\delta\cI^\three_{ji}(x,b_\perp,&\,\mu,\nu) = \Lp^3\bigg[
\left(\frac{1}{2}\beta_0+\frac{1}{4}(2 \gamma^B_0-\Gcusp_0 L_Q) \right) \sum_{l}{\delta P}^\zero_{jl} \otimes {\delta P}^\zero_{li}(x) 
\nn\\
-&\frac{1}{6} \sum_{l\,k } {\delta P}^\zero_{jl} \otimes {\delta P}^\zero_{lk} \otimes {\delta P}^\zero_{ki}(x) 
+\delta_{ji} \delta(1-x) \left( \frac{1}{6}\beta_0^2(2 \gamma^B_0-\Gcusp_0 L_Q)\right.
\nn\\
+&\left.\frac{1}{8} \beta_0(2 \gamma^B_0-\Gcusp_0 L_Q)^2+\frac{1}{48} (2 \gamma^B_0-\Gcusp_0 L_Q)^3 \right) \nn
\\
+&{\delta P}^\zero_{ji}  \left( -\frac{1}{2} \beta_0 (2 \gamma^B_0-\Gcusp_0 L_Q)-\frac{1}{3}\beta_0^2-\frac{1}{8} (2 \gamma^B_0-\Gcusp_0 L_Q)^2 \right) 
\bigg]
\nn\\
+&\Lp^2\bigg[ 
\left(-\frac{3}{2} \beta_0-\frac{1}{2}(2 \gamma^B_0-\Gcusp_0 L_Q) \right)\sum_l \delta I^\one_{jl} \otimes {\delta P}^\zero_{li} (x) 
\nn\\
+&\frac{1}{2} \sum_{l\,k } \delta I^\one_{jl} \otimes {\delta P}^\zero_{lk} \otimes {\delta P}^\zero_{ki}(x)
+\frac{1}{2} \sum_l {\delta P}^\zero_{jl} \otimes {\delta P}^\one_{li}(x)
\nn\\
+&{\delta P}^\zero_{ji}(x) \left(-\frac{1}{2}\beta_1-\frac{1}{2}(2 \gamma^B_1-\Gcusp_1 L_Q) \right)
+\delta_{ji} \delta(1-x)  \left ( \frac{1}{4} \beta_1(2 \gamma^B_0-\Gcusp_0 L_Q)\right.
\nn\\
+&\left.\frac{1}{2}\beta_0(2 \gamma^B_1-\Gcusp_1 L_Q)+\frac{1}{4}(2 \gamma^B_0-\Gcusp_0 L_Q)(2 \gamma^B_1-\Gcusp_1 L_Q)
 \right)  \nn
\\
+&\delta I^\one_{ji}(x) \left( \frac{3}{4} \beta_0(2 \gamma^B_0-\Gcusp_0 L_Q)+\beta_0^2+\frac{1}{8}(2 \gamma^B_0-\Gcusp_0 L_Q)^2  \right)
\nn\\
+&{\delta P}^\one_{ji}(x) \left(-\beta_0-\frac{1}{2}(2 \gamma^B_0-\Gcusp_0 L_Q) \right)
+\frac{1}{2}\sum_l {\delta P}^\one_{jl} \otimes {\delta P}^\zero_{li}(x) 
\bigg] \nn
\\
+&\Lp \bigg[
-\sum_{l }\delta I^\one_{jl} \otimes {\delta P}^\one_{li} (x)-\sum_{l }\delta I^\two_{jl} \otimes {\delta P}^\zero_{li} (x) -{\delta P}^\zero_{ji}(x) \gamma^R_1 L_Q
-{\delta P}^\two_{ji}(x) 
\nn\\
+&\delta_{ji} \delta(1-x)  \biggl( 2 \beta_0  \gamma_1^R L_Q 
 +\frac{1}{2}\gamma_1^R (2 \gamma^B_0-\Gcusp_0 L_Q)L_Q+\frac{1}{2} (2\gamma_2^B-\Gcusp_2 L_Q) \biggl )
 \nn\\
 +&\delta I^\one_{ji} (x)\left( \beta_1+\frac{1}{2}(2 \gamma^B_1-\Gcusp_1 L_Q) \right) \nn
+\delta I^\two_{ji} (x)\left(2 \beta_0+\frac{1}{2}(2 \gamma^B_0-\Gcusp_0 L_Q) \right)\bigg]
\nn\\
+&\delta_{ji} \delta(1-x) \gamma^R_2 L_Q+\delta I^\one_{ji}(x) \gamma^R_1 L_Q+ \delta I^\three_{ji} (x)\,,
\end{align}
where $\delta I^{(n)}_{ji} (x)$ are the scale-independent coefficient functions, whose small-$x$ asymptotic behavior is collected in Sec.~(\ref{sec:small-x-expansion-PDF}),
 while numerical fits are provided in Sec.~(\ref{sec:num-I}).
We have kept the dependence on $\gamma_0^R$  explicit in the general solution, despite  the fact that $\gamma_0^R = 0$.
 However, in the expression for $\delta\cI_{ji}^{\three}$, the condition   $\gamma_0^R = 0$ has been used to simplify the result.
 The scale logarithms are defined by
\begin{align}
\label{eq:LdefinitionS}
 L_\perp = \ln \frac{b_T^2 \mu^2}{b_0^2} , \quad  L_Q  = 2 \ln \frac{x \,  P_+}{\nu}, \quad L_\nu = \ln \frac{\nu^2}{\mu^2} \,,\quad b_0 =2  e^{- \gamma_E}\,.
\end{align}
Similarly, the solution to the fragmentation coefficient functions are
\begin{align}
\label{eq:RGs-2}
\delta\cC^\zero_{ji}(z,b_\perp/z,&\,\mu,\nu) =  \delta_{ji} \delta(1-z) \, ,\nn
\\
\delta\cC^\one_{ji}(z,b_\perp/z,&\,\mu,\nu) = \left( - \frac{\Gcusp_0}{2} \Lp L_Q + \gamma_0^B \Lp + \gamma_0^R L_Q \right) \delta_{ji} \delta(1-z) - {\delta P}_{ji}^{T\zero}(z) \Lp 
\nn\\
+&\delta C_{ji}^\one(z) \, , \nn
\\
\delta\cC^\two_{ji}(z,b_\perp/z,&\,\mu,\nu) =  \bigg[ \frac{1}{8} \left( -\Gcusp_0 L_Q + 2\gamma^B_0 \right) \left( -\Gcusp_0 L_Q + 2\gamma^B_0 + 2\beta_0 \right) \Lp^2
\nn\\
+& \left(  (-\Gcusp_0 L_Q + 2\gamma^B_0 + 2\beta_0) \frac{\gamma_0^R}{2} L_Q 
-\frac{\Gcusp_1}{2} L_Q + \gamma^B_1  \right) \Lp 
\nn\\
+& \frac{(\gamma_0^R)^2}{2} L_Q^2 + \gamma_1^R L_Q \bigg] \, \delta_{ji} \delta(1-z) 
+ \bigg( \frac{1}{2} \sum_l {\delta P}^{T\zero}_{jl} \otimes {\delta P}^{T\zero}_{li}(z) \nn
\\
+& \frac{{\delta P}^{T\zero}_{ji}(z)}{2} (\Gcusp_0 L_Q - 2\gamma_0^B - \beta_0) \bigg) \Lp^2 + \bigg[ -{\delta P}^{T\one}_{ji}(z) - {\delta P}^{T\zero}_{ji}(z) \gamma_0^R L_Q 
\nn\\
- &\sum_l {\delta P}^{T\zero}_{jl} \otimes \delta C^\one_{li}(z) \nn
+ \left( -\frac{\Gcusp_0}{2} L_Q + \gamma_0^B + \beta_0 \right)\delta C^\one_{ji}(z) \bigg] \Lp
\nn\\
 +& \gamma_0^R L_Q \delta C^\one_{ji}(z) + \delta C^\two_{ji}(z) \, ,\nn
\\
\delta\cC^\three_{ji}(z,b_\perp/z,&\,\mu,\nu) = \Lp^3\bigg[
\left(\frac{1}{2}\beta_0+\frac{1}{4}(2 \gamma^B_0-\Gcusp_0 L_Q) \right) \sum_{l}{\delta P}^{T\zero}_{jl} \otimes {\delta P}^{T\zero}_{li}(z) 
\nn\\
-&\frac{1}{6} \sum_{l\,k } {\delta P}^{T\zero}_{jl} \otimes {\delta P}^{T\zero}_{lk} \otimes {\delta P}^{T\zero}_{ki}(z) 
+\delta_{ji} \delta(1-z) \left( \frac{1}{6}\beta_0^2(2 \gamma^B_0-\Gcusp_0 L_Q)\right.
\nn\\
+&\left.\frac{1}{8} \beta_0(2 \gamma^B_0-\Gcusp_0 L_Q)^2+\frac{1}{48} (2 \gamma^B_0-\Gcusp_0 L_Q)^3 \right)
 \nn\\
+&{\delta P}^{T\zero}_{ji}  \left( -\frac{1}{2} \beta_0 (2 \gamma^B_0-\Gcusp_0 L_Q)-\frac{1}{3}\beta_0^2-\frac{1}{8} (2 \gamma^B_0-\Gcusp_0 L_Q)^2 \right) 
\bigg]
\nn\\
+&\Lp^2\bigg[ 
\left(-\frac{3}{2} \beta_0-\frac{1}{2}(2 \gamma^B_0-\Gcusp_0 L_Q) \right)\sum_l {\delta P}^{T\zero}_{jl} \otimes \delta C^\one_{li} (z) 
\nn\\
+&\frac{1}{2} \sum_{l\,k } {\delta P}^{T\zero}_{jl} \otimes {\delta P}^{T\zero}_{lk} \otimes\delta  C^\one_{ki}(z)
+\frac{1}{2} \sum_l {\delta P}^{T\zero}_{jl} \otimes {\delta P}^{T\one}_{li}(z)
\nn\\
+&{\delta P}^{T\zero}_{ji}(z) \left(-\frac{1}{2}\beta_1-\frac{1}{2}(2 \gamma^B_1-\Gcusp_1 L_Q) \right)
+\delta_{ji} \delta(1-z)  \left ( \frac{1}{4} \beta_1(2 \gamma^B_0-\Gcusp_0 L_Q)\right.
\nn\\
+&\left.\frac{1}{2}\beta_0(2 \gamma^B_1-\Gcusp_1 L_Q)+\frac{1}{4}(2 \gamma^B_0-\Gcusp_0 L_Q)(2 \gamma^B_1-\Gcusp_1 L_Q)
 \right)  
 \nn\\
+&\delta C^\one_{ji}(z) \left( \frac{3}{4} \beta_0(2 \gamma^B_0-\Gcusp_0 L_Q)+\beta_0^2+\frac{1}{8}(2 \gamma^B_0-\Gcusp_0 L_Q)^2  \right)
\nn\\
+&{\delta P}^{T\one}_{ji}(z) \left(-\beta_0-\frac{1}{2}(2 \gamma^B_0-\Gcusp_0 L_Q) \right)
+\frac{1}{2}\sum_l {\delta P}^{T\one}_{jl} \otimes {\delta P}^{T\zero}_{li}(z) 
\bigg] 
\nn\\
+&\Lp \bigg[
-\sum_{l } {\delta P}^{T\one}_{jl} \otimes\delta C^\one_{li} (z)-\sum_{l } {\delta P}^{T\zero}_{jl} \otimes \delta C^\two_{li} (z) -{\delta P}^{T\zero}_{ji}(z) \gamma^R_1 L_Q
\nn\\
-&{\delta P}^{T\two}_{ji}(z) 
+\delta_{ji} \delta(1-z)  \biggl( 2 \beta_0  \gamma_1^R L_Q 
 +\frac{1}{2}\gamma_1^R (2 \gamma^B_0-\Gcusp_0 L_Q)L_Q+\frac{1}{2} (2\gamma_2^B
  \nn\\
 -&\Gcusp_2 L_Q) \biggl )
 +\delta C^\one_{ji} (z)\left( \beta_1+\frac{1}{2}(2 \gamma^B_1-\Gcusp_1 L_Q) \right) 
 \nn\\
+&\delta C^\two_{ji} (z)
\left(2 \beta_0
+\frac{1}{2}(2 \gamma^B_0-\Gcusp_0 L_Q) 
\right)\bigg]
+\delta_{ji} \delta(1-z) \gamma^R_2 L_Q+\delta C^\one_{ji}(z) \gamma^R_1 L_Q
\nn\\
+&\delta C^\three_{ji} (z)\,.
\end{align}
where $\delta C^{(n)}_{ji} (z)$ are the scale-independent coefficient functions, whose small-$z$ asymptotic behavior is  collected in Sec.~(\ref{sec:small-x-expansion-FF}),
 while numerical fits are provided in Sec.~(\ref{sec:num-C}).
Again, we have kept the dependence on $\gamma_0^R$  explicit in the general solution, despite  the fact that $\gamma_0^R = 0$.
 However, in the expression for $\delta\cC^\three_{ji}$, the condition   $\gamma_0^R = 0$ has been used to simplify the result.
The anomalous dimensions appeared above are identical to those in the space-like case.
The logarithms appeared in the fragmentation coefficient functions are defined as
\begin{align}
\label{eq:LdefinitionT}
 L_\perp = \ln \frac{b_T^2 \mu^2}{b_0^2} , \quad L_Q = 2 \ln \frac{ P_+}{ z \, \nu}, \quad L_\nu = \ln \frac{\nu^2}{\mu^2} \,,\quad b_0 =2  e^{- \gamma_E}\,.
\end{align}
Both space-like and time-like coefficient functions depend on the rapidity regulator being used. 
Rapidity-regulator-independent TMD transversity PDFs and FFs can be obtained by 
multiplying the coefficient functions with the square root of the TMD soft functions ${\cal S}(b_\perp, \mu, \nu)$~\cite{Luo:2019hmp,Luo:2019bmw}.
As a result, the rapidity cut-off dependence in the collinear beam functions cancels against that of the TMD soft functions, leaving a physical rapidity scale 
$\xi$ in the TMDs, with $\sqrt{\xi}=x P_+$ for TMD PDFs and $\sqrt{\xi}= P_+/z$ for TMD FFs.
Furthermore, the following physical hard-collinear contributions can be defined in terms of the physical rapidity logarithm $L_h=L_Q+L_\nu=\ln(\xi/\mu^2)$
\begin{align}
  \label{eq:TMD- PDF-FF}
 \delta \mathscr{I}_{qj}(x, b_\perp, \mu,\xi) =&\delta {\cal I}_{qj}(x, b_\perp, \mu, \nu) \sqrt{{\cal S}(b_\perp, \mu, \nu)} \,,
\nn\\
\delta \mathscr{C}_{iq}(z, b_\perp/z,  \mu,\xi) = &\delta{\cal C}_{iq}(z, b_\perp/z,  \mu, \nu) \sqrt{{\cal S}(b_\perp, \mu, \nu)} \,.
\end{align}

 \section{The NNLO  transversity splitting functions }
\label{sec:n3lo-sp-transversity}
The next-to-next-to-leading order (NNLO) space-like transversity splitting functions with massless quarks were first obtained in Refs.~\cite{Vogelsang:1997ak,Mikhailov:2008my,Blumlein:2021enk},
contributions from  heavy quark flavors   were  considered in Refs.~\cite{Blumlein:2009rg,Ablinger:2014vwa}.
In this work, we extend the precision also for the time-like ones to next-to-next-to-leading order (NNLO) for the first time. A detailed comparison with the known NNLO results confirms complete agreement. In the following, we present our analytical results for both the space-like and time-like transversity splitting functions at NNLO accuracy.

 \subsection{Space-like results in ${\overline{\text{MS}}}$}
 \begin{dmath}[style={\small},compact]
  \delta P_{qq}^{(0)}=
  4 C_F \left[\frac{1}{1-x}\right]_++3 C_F \delta (1-x)-4 C_F
 \,,
\end{dmath}
 \begin{dmath}[style={\small},compact]
  \delta P_{qq}^{(1)}=
  \left[\frac{1}{1-x}\right]_+ \left(\left(\frac{268}{9}-8 \zeta _2\right) C_A C_F-\frac{40 C_F N_f}{9}\right)+C_A C_F \left(\frac{x}{x-1}\left(\left(-\frac{44}{3}\right)H_0-8 H_{0,0}\right)+2 x+8 \zeta _2-\frac{286}{9}\right)+C_F^2 \left(\frac{x}{x-1}\left(-16 H_{1,0}+12 H_0-16 H_2\right)-4 x+4\right)+\delta (1-x) \left(\left(\frac{44}{3}\zeta _2-12 \zeta _3+\frac{17}{6}\right) C_A C_F+\left(\left(-\frac{8}{3}\right)\zeta _2-\frac{1}{3}\right) C_F N_f+\left(-12 \zeta _2+24 \zeta _3+\frac{3}{2}\right) C_F^2\right)+C_F N_f T_F \left(\frac{16 x}{3 (x-1)}H_0+\frac{80}{9}\right)
 \,,
\end{dmath}
 \begin{dmath*}[style={\small},compact]
  \delta P_{qq}^{(2)}=
        C_A^2 C_F \left(4 H_{-1} \zeta _2 x^2+\left(\left(-\frac{4}{3}\right)H_{-1,0}-8 H_{-1,2}-8 H_{-1,-1,0}\right) x^2+36 H_{-1} \zeta _2 x+\left(8 H_2+4 H_{-1,0}-72 H_{-1,2}-72 H_{-1,-1,0}\right) x+\frac{1}{x-1}\left(\left(-\frac{4}{3}\right)H_0+\frac{92}{3}H_1+\frac{1586}{9}\right)+\frac{88}{3 (x-1)}\zeta _3+\frac{88}{x-1}\zeta _4+\frac{1}{x}\left(\left(-\frac{4}{3}\right)H_{-1,0}-8 H_{-1,2}-8 H_{-1,-1,0}\right)+\left(-\frac{192 x}{x-1}\right)\zeta _4+\frac{x}{x-1}\left(\left(-\frac{5396}{27}\right)H_0+\left(-\frac{1592}{9}\right)H_{0,0}+\left(-\frac{304}{3}\right)H_3+\left(-\frac{248}{3}\right)H_{0,0,0}+\left(-\frac{184}{3}\right)H_1-32 H_4-32 H_{-3,0}-24 H_{-2,0}-64 H_{-2,2}-128 H_{1,3}-128 H_{-2,-1,0}+32 H_{-2,0,0}-192 H_{1,-2,0}+88 H_{1,0,0}+64 H_{2,0,0}+32 H_{0,0,0,0}+96 H_{1,0,0,0}+128 H_{1,1,0,0}-\frac{1702}{9}\right)+\left(-\frac{32 x^2}{x-1}\right)\zeta _3+\frac{x^2}{x-1}\left(\frac{8}{3}H_{0,0}+\frac{92}{3}H_1+\frac{140}{3}H_0+64 H_3+64 H_{-2,0}+\frac{116}{9}\right)+\left(-\frac{4 x^3}{x-1}\right)\zeta _3+\frac{x^3}{x-1}\left(\frac{4}{3}H_{0,0}+8 H_3+8 H_{-2,0}\right)+\frac{4 \zeta _2}{x}H_{-1}+\left(-\frac{4 \zeta _2}{(x-1) x}\right)H_1+\frac{1}{x-1}\left(-32 H_1-\frac{1000}{9}\right)\zeta _2+\frac{x}{x-1}\left(-128 H_0-288 H_1+20\right)\zeta _3
      +\frac{x}{x-1}\left(\frac{464 H_0}{3}+72 H_1+32 H_{1,0}+\frac{1036}{9}\right)\zeta _2+\frac{x^2}{x-1}\left(-64 H_0-32 H_1-\frac{8}{3}\right)\zeta _2
      +\frac{x^3}{x-1}\left(-8 H_0-4 H_1-\frac{4}{3}\right)\zeta _2+8 H_2+36 H_{-1} \zeta _2+4 H_{-1,0}-72 H_{-1,2}-72 H_{-1,-1,0}\right) 
           +
    N_f T_F C_F^2\bigg(\frac{64}{x-1}\zeta _3+\left(-\frac{320 x}{3 (x-1)}\right)\zeta _3+\frac{x}{x-1}\left(\frac{64}{3}H_{0,0,0}+\frac{128}{3}H_3+\frac{128}{3}H_{2,0}+\frac{640}{9}H_2+\frac{640}{9}H_{1,0}+\frac{256}{3}H_{1,0,0}-24 H_0-32 H_{0,0}+\frac{436}{9}\right)
    \end{dmath*}
 \begin{dmath*}[style={\small},compact]
 +\frac{x^2}{x-1}\left(\frac{32}{3}H_0+\frac{112}{9}\right)+\left(-\frac{64 x \zeta _2}{3 (x-1)}\right)H_0-\frac{548}{9 (x-1)}\bigg) 
    +N_f^2 T_F^2 C_F\left(-\frac{32 x}{9}+\frac{x}{x-1}\left(\left(-\frac{320}{27}\right)H_0+\left(-\frac{64}{9}\right)H_{0,0}\right)+\frac{160}{27}\right)
      +C_F^3 \left(\left(\left(-\frac{64}{3}\right)H_{-1,-1,0}+\left(-\frac{32}{3}\right)H_{-1,2}+\left(-\frac{16}{3}\right)H_{0,0,0}+\frac{8}{3}H_{-1,0}+\frac{16}{3}H_{-1,0,0}+\frac{32}{3}H_3+\frac{64}{3}H_{-2,0}\right) x^2+\left(\left(-\frac{16}{3}\right)H_2+\frac{40}{3}H_{-1,0}+96 H_3+192 H_{-2,0}-96 H_{-1,2}-16 H_{1,0}-192 H_{-1,-1,0}+48 H_{-1,0,0}-48 H_{0,0,0}\right) x+\frac{40}{3}H_{-1,0}+\frac{80}{3}H_2+\frac{1}{x-1}\left(\left(-\frac{8}{3}\right)H_0+\frac{16}{3}H_{0,0}+\frac{200}{3}H_1+\frac{104}{3}\right)+\frac{1}{x}\left(\left(-\frac{64}{3}\right)H_{-1,-1,0}+\left(-\frac{32}{3}\right)H_{-1,2}+\frac{8}{3}H_{-1,0}+\frac{16}{3}H_{-1,0,0}\right)+\left(-\frac{72 x}{x-1}\right)\zeta _4+\frac{x}{x-1}\left(\left(-\frac{400}{3}\right)H_1+\left(-\frac{254}{3}\right)H_0-64 H_4+64 H_{-3,0}-28 H_{0,0}-256 H_{1,3}+96 H_{2,0}-128 H_{2,2}-128 H_{3,0}-128 H_{3,1}-384 H_{-2,-1,0}+192 H_{-2,0,0}-256 H_{1,-2,0}+192 H_{1,0,0}-128 H_{1,2,0}-64 H_{2,0,0}-128 H_{2,1,0}+64 H_{0,0,0,0}+128 H_{1,0,0,0}-\frac{208}{3}\right)+\frac{x^2}{x-1}\left(\left(-\frac{80}{3}\right)H_{0,0}+\frac{200}{3}H_1+\frac{280}{3}H_0+\frac{104}{3}\right)+\frac{128 x^2}{3 (x-1)}\zeta _3+\left(-\frac{8 x^3}{3 (x-1)}\right)H_{0,0}+\frac{16 x^3}{3 (x-1)}\zeta _3+\left(-\frac{16 \zeta _2}{3 (x-1) x}\right)H_1+\frac{1}{x-1}\left(\frac{64}{3}-\frac{128 H_1}{3}\right)\zeta _2+\frac{x}{x-1}\left(-32 H_0-384 H_1-48\right)\zeta _3+\frac{x}{x-1}\left(-192 H_{-2}+96 H_0+96 H_1-\frac{104}{3}\right)\zeta _2+\frac{x^2}{x-1}\left(-\frac{128 H_0}{3}-\frac{128 H_1}{3}+\frac{32}{3}\right)\zeta _2+\frac{x^3}{x-1}\left(-\frac{16 H_0}{3}-\frac{16 H_1}{3}+\frac{8}{3}\right)\zeta _2-96 H_{-1,2}+16 H_{1,0}-192 H_{-1,-1,0}+48 H_{-1,0,0}\right) 
  +C_A C_F^2\bigg(\left(\left(-\frac{8}{3}\right)H_{-1,0,0}+\frac{4}{3}H_{-1,0}+\frac{64}{3}H_{-1,2}+\frac{80}{3}H_{-1,-1,0}\right) x^2-8 H_{-1} \zeta _2 x^2-72 H_{-1} \zeta _2 x+\left(\left(-\frac{44}{3}\right)H_{-1,0}+192 H_{-1,2}+240 H_{-1,-1,0}-24 H_{-1,0,0}\right) x+\left(-\frac{44}{3}\right)H_{-1,0}+\frac{1}{x-1}\left(\left(-\frac{328}{3}\right)H_1+\left(-\frac{8}{3}\right)H_{0,0}+\frac{88}{3}H_2+4 H_0+8 H_{1,0}-\frac{872}{9}\right)+\frac{1}{x}\left(\left(-\frac{8}{3}\right)H_{-1,0,0}+\frac{4}{3}H_{-1,0}+\frac{64}{3}H_{-1,2}+\frac{80}{3}H_{-1,-1,0}\right)
     +\frac{x}{x-1}\left(\left(-\frac{992}{3}\right)H_{1,0,0}+\left(-\frac{2288}{9}\right)H_2+\left(-\frac{2288}{9}\right)H_{1,0}+\left(-\frac{352}{3}\right)H_{2,0}+\frac{112}{3}H_{0,0,0}+\frac{512}{3}H_3+\frac{656}{3}H_1+\frac{754}{3}H_0-32 H_4+32 H_{-3,0}+144 H_{-2,0}+128 H_{-2,2}+100 H_{0,0}+256 H_{1,3}-32 H_{3,0}+448 H_{-2,-1,0}-160 H_{-2,0,0}+512 H_{1,-2,0}-160 H_{2,0,0}-160 H_{0,0,0,0}-352 H_{1,0,0,0}-256 H_{1,1,0,0}+\frac{1744}{9}\right)
        +\frac{276 x}{x-1}\zeta _4+\frac{x^2}{x-1}\left(\left(-\frac{640}{3}\right)H_{-2,0}+\left(-\frac{512}{3}\right)H_3+\left(-\frac{464}{3}\right)H_0+\left(-\frac{328}{3}\right)H_1+\left(-\frac{40}{3}\right)H_2+\frac{64}{3}H_{0,0,0}+8 H_{0,0}+8 H_{1,0}-\frac{872}{9}\right)
        +\frac{128 x^2}{3 (x-1)}\zeta _3+\frac{x^3}{x-1}\left(\left(-\frac{80}{3}\right)H_{-2,0}+\left(-\frac{64}{3}\right)H_3+\left(-\frac{4}{3}\right)H_{0,0}+\frac{8}{3}H_{0,0,0}\right)
        +\frac{16 x^3}{3 (x-1)}\zeta _3+\left(-\frac{8 \zeta _2}{x}\right)H_{-1}+\frac{32 \zeta _2}{3 (x-1) x}H_1
        +\frac{1}{x-1}\left(\frac{256 H_1}{3}-\frac{80}{3}\right)\zeta _2+\frac{x}{x-1}\left(272 H_0+768 H_1-\frac{224}{3}\right)\zeta _3
        +\frac{x}{x-1}\left(96 H_{-2}-\frac{832 H_0}{3}-192 H_1+64 H_2+128 H_{0,0}+64 H_{1,0}+\frac{76}{3}\right)\zeta _2
        +\frac{x^2}{x-1}\left(\frac{448 H_0}{3}+\frac{256 H_1}{3}\right)\zeta _2
           +\frac{x^3}{x-1}\left(\frac{56 H_0}{3}+\frac{32 H_1}{3}+\frac{4}{3}\right)\zeta _2-72 H_{-1} \zeta _2+192 H_{-1,2}+240 H_{-1,-1,0}-24 H_{-1,0,0}\bigg)  
        \end{dmath*}
 \begin{dmath}[style={\small},compact]
   +\delta (1-x) \left(\left(-32 \zeta _3 \zeta _2+18 \zeta _2+68 \zeta _3+144 \zeta _4-240 \zeta _5+\frac{29}{2}\right) C_F^3+\left(\left(-\frac{136}{3}\right)\zeta _3+\frac{20}{3}\zeta _2+\frac{116}{3}\zeta _4-23\right) N_f C_F^2+\left(\left(-\frac{16}{9}\right)\zeta _3+\frac{80}{27}\zeta _2-\frac{17}{9}\right) N_f^2 C_F+C_A^2 \left(\left(-\frac{1552}{9}\right)\zeta _3+\frac{4496}{27}\zeta _2-5 \zeta _4+40 \zeta _5-\frac{1657}{36}\right) C_F+C_A \left(\left(\left(-\frac{494}{3}\right)\zeta _4+\left(-\frac{410}{3}\right)\zeta _2+\frac{844}{3}\zeta _3+16 \zeta _2 \zeta _3+120 \zeta _5+\frac{151}{4}\right) C_F^2+N_f \left(\left(-\frac{1336}{27}\right)\zeta _2+\frac{200}{9}\zeta _3+2 \zeta _4+20\right) C_F\right)\right)
  +C_A  N_f T_F C_F\left(\left(-\frac{8}{3}\right)H_1+\left(-\frac{224}{3 (x-1)}\right)\zeta _3+\frac{320}{9 (x-1)}\zeta _2+\frac{8 x}{3}H_1+\frac{x}{x-1}\left(\frac{32}{3}H_3+\frac{64}{3}H_{0,0,0}+\frac{224}{3}H_{0,0}+\frac{2744}{27}H_0-32 H_{1,0,0}+\frac{952}{27}\right)+\frac{96 x}{x-1}\zeta _3+\frac{x^2}{x-1}\left(\left(-\frac{8}{3}\right)H_0+\frac{40}{3}\right)+\frac{x}{x-1}\left(-\frac{64 H_0}{3}-\frac{320}{9}\right)\zeta _2-\frac{1312}{27 (x-1)}\right) 
  +\left[\frac{1}{1-x}\right]_+ \left(C_F \left(\left(-\frac{1072}{9}\right)\zeta _2+88 \zeta _4+\frac{88 \zeta_3}{3}+\frac{490}{3}\right) C_A^2+C_F N_f \left(\frac{160}{9}\zeta _2-\frac{112 \zeta_3}{3}-\frac{836}{27}\right) C_A-\frac{16}{27} C_F N_f^2+C_F^2 N_f \left(-\frac{110}{3}+32 \zeta_3\right)\right)
 \,,
\end{dmath}
 \begin{dmath}[style={\small},compact]
  \delta P_{q\bar q}^{(1)}=
  C_A C_F \left(\frac{x}{x+1}\left(8 H_{0,0}-16 H_{-1,0}\right)-\frac{2 x^2}{x+1}+\left(-\frac{8 x}{x+1}\right)\zeta _2+\frac{2}{x+1}\right)+C_F^2 \left(\frac{x}{x+1}\left(32 H_{-1,0}-16 H_{0,0}\right)+\frac{4 x^2}{x+1}+\frac{16 x}{x+1}\zeta _2-\frac{4}{x+1}\right)
 \,,
\end{dmath}
\begin{dmath*}[style={\small},compact]
 \delta P_{q\bar q}^{(2)}=
  N_f T_F C_F C_A\left(\frac{56 x^2}{9 (x+1)}+\frac{16}{3}H_1+\left(-\frac{16 x}{3}\right)H_1+\left(-\frac{32 x}{x+1}\right)\zeta _3+\frac{x}{x+1}\left(\left(-\frac{128}{3}\right)H_{-1,2}+\left(-\frac{320}{9}\right)H_{0,0}+\left(-\frac{64}{3}\right)H_{0,0,0}+\frac{64}{3}H_3+\frac{64}{3}H_{-2,0}+\frac{64}{3}H_{-1,0,0}+\frac{640}{9}H_{-1,0}\right)+\frac{x}{x+1}\left(\frac{128 H_{-1}}{3}-\frac{32 H_0}{3}+\frac{320}{9}\right)\zeta _2-\frac{56}{9 (x+1)}\right) 
   +N_f T_F C_F^2\left(\frac{x}{x+1}\left(\left(-\frac{1280}{9}\right)H_{-1,0}
   +\left(-\frac{128}{3}\right)H_3+\left(-\frac{128}{3}\right)H_{-2,0}+\left(-\frac{128}{3}\right)H_{-1,0,0}+\frac{128}{3}H_{0,0,0}+\frac{640}{9}H_{0,0}+\frac{256}{3}H_{-1,2}\right)
   -\frac{112 x^2}{9 (x+1)}+\left(-\frac{32}{3}\right)H_1+\frac{32 x}{3}H_1+
   +\frac{64 x}{x+1}\zeta _3
+\frac{x}{x+1}\left(-\frac{256 H_{-1}}{3}+\frac{64 H_0}{3}-\frac{640}{9}\right)\zeta _2
+\frac{112}{9 (x+1)}\right) 
+\frac{1}{x+1}\left(128 H_{-1}+\frac{64}{3}\right)\zeta _2+\frac{x}{x+1}\left(1344 H_{-1}-464 H_0-416\right)\zeta _3+\frac{x}{x+1}\left(1024 H_{-2}+\frac{1856 H_{-1}}{3}-\frac{896 H_0}{3}+224 H_2-1792 H_{-1,-1}+1344 H_{-1,0}-416 H_{0,0}+\frac{2540}{9}\right)\zeta _2+\frac{x^2}{x+1}\left(128 H_{-1}-\frac{704 H_0}{3}+\frac{64}{3}\right)\zeta _2+\frac{x^3}{x+1}\left(-16 H_{-1}+\frac{88 H_0}{3}-\frac{4}{3}\right)\zeta _2+76 H_1-120 x H_1 \zeta _2+120 H_1 \zeta _2+8 H_{1,0}+x \left(-76 H_1-8 H_{1,0}-192 H_{-1,-1,0}\right)-192 H_{-1,-1,0}
      \end{dmath*}
\begin{dmath}[style={\small},compact]
+
C_F^3  \left(\left(-\frac{184}{3}\right)H_1+\left(-\frac{32}{3 x}\right)H_{-1,-1,0}+\left(-\frac{32 x^2}{3}\right)H_{-1,-1,0}+\frac{1}{x+1}\left(\left(-\frac{8}{3}\right)H_0+\frac{16}{3}H_2+\frac{16}{3}H_{0,0}+\frac{128}{3}H_{-1,0,0}+\frac{512}{3}H_{-1,2}+\frac{104}{3}\right)
+\frac{1}{x (x+1)}\left(\left(-\frac{64}{3}\right)H_{-1,2}+\left(-\frac{16}{3}\right)H_{-1,0,0}+\left(-\frac{8}{3}\right)H_{-1,0}\right)+\left(-\frac{280 x}{x+1}\right)\zeta _4
+\frac{x}{x+1}\left(\frac{16}{3}H_{-1,0}+\frac{128}{3}H_2+\frac{152}{3}H_{0,0}+80 H_0-288 H_3-384 H_4+192 H_{-3,0}-192 H_{-2,0}+896 H_{-2,2}+576 H_{-1,2}+1024 H_{-1,3}-64 H_{3,0}-256 H_{-2,-1,0}+832 H_{-2,0,0}-256 H_{-1,-2,0}-1536 H_{-1,-1,2}+128 H_{-1,2,0}+96 H_{0,0,0}-1280 H_{-1,-1,0,0}+704 H_{-1,0,0,0}-192 H_{0,0,0,0}\right)+\frac{x^2}{x+1}\left(\left(-\frac{512}{3}\right)H_3+\left(-\frac{256}{3}\right)H_{-2,0}+\left(-\frac{128}{3}\right)H_{0,0,0}+\frac{104}{3}H_0+\frac{112}{3}H_2+\frac{128}{3}H_{-1,0,0}+\frac{512}{3}H_{-1,2}+48 H_{0,0}-\frac{104}{3}\right)+\frac{512 x^2}{3 (x+1)}\zeta _3+\left(-\frac{64 x^3}{3 (x+1)}\right)\zeta _3+\frac{x^3}{x+1}\left(\left(-\frac{64}{3}\right)H_{-1,2}+\left(-\frac{16}{3}\right)H_{-1,0,0}+\left(-\frac{8}{3}\right)H_{-1,0}+\frac{8}{3}H_{0,0}+\frac{16}{3}H_{0,0,0}+\frac{32}{3}H_{-2,0}+\frac{64}{3}H_3\right)+\frac{32 \zeta _2}{3 x}H_1+\left(-\frac{32}{3} x^2 \zeta _2\right)H_1+\frac{16 \zeta _2}{x (x+1)}H_{-1}+\frac{1}{x+1}\left(-128 H_{-1}-\frac{32}{3}\right)\zeta _2+\frac{x}{x+1}\left(-1152 H_{-1}+416 H_0+336\right)\zeta _3+\frac{x}{x+1}\left(-1024 H_{-2}-480 H_{-1}+288 H_0-192 H_2+1536 H_{-1,-1}-1280 H_{-1,0}+448 H_{0,0}-40\right)\zeta _2+\frac{x^2}{x+1}\left(-128 H_{-1}+\frac{640 H_0}{3}-32\right)\zeta _2+\frac{x^3}{x+1}\left(16 H_{-1}-\frac{80 H_0}{3}-\frac{8}{3}\right)\zeta _2+96 x H_1 \zeta _2-96 H_1 \zeta _2-16 H_{1,0}+96 H_{-1,-1,0}+x \left(\frac{184}{3}H_1+16 H_{1,0}+96 H_{-1,-1,0}\right)\right) 
  + C_A C_F^2\left(\frac{64}{3 x}H_{-1,-1,0}+\frac{64 x^2}{3}H_{-1,-1,0}+\frac{1}{x+1}\left(\left(-\frac{640}{3}\right)H_{-1,2}+\left(-\frac{64}{3}\right)H_{-1,0,0}+\left(-\frac{56}{3}\right)H_2+\left(-\frac{32}{3}\right)H_{-1,0}+\left(-\frac{8}{3}\right)H_{0,0}+4 H_0-\frac{872}{9}\right)+\frac{1}{x (x+1)}\left(\left(-\frac{4}{3}\right)H_{-1,0}+\frac{8}{3}H_{-1,0,0}+\frac{80}{3}H_{-1,2}\right)+\left(-\frac{20 x}{x+1}\right)\zeta _4+\frac{x}{x+1}\left(\left(-\frac{2432}{3}\right)H_{-1,2}+\left(-\frac{2444}{9}\right)H_{0,0}+\left(-\frac{640}{3}\right)H_{0,0,0}+\left(-\frac{296}{3}\right)H_0+\left(-\frac{160}{3}\right)H_2+\frac{352}{3}H_{-1,0,0}+\frac{1072}{3}H_{-2,0}+\frac{1216}{3}H_3+\frac{4120}{9}H_{-1,0}+320 H_4-160 H_{-3,0}-960 H_{-2,2}-1024 H_{-1,3}+32 H_{3,0}+128 H_{-2,-1,0}-672 H_{-2,0,0}+128 H_{-1,-2,0}+1792 H_{-1,-1,2}-64 H_{-1,2,0}+1152 H_{-1,-1,0,0}-544 H_{-1,0,0,0}+160 H_{0,0,0,0}\right)+\left(-\frac{448 x^2}{3 (x+1)}\right)\zeta _3+\frac{x^2}{x+1}\left(\left(-\frac{640}{3}\right)H_{-1,2}+\left(-\frac{104}{3}\right)H_2+\left(-\frac{92}{3}\right)H_0+\left(-\frac{88}{3}\right)H_{0,0}+\left(-\frac{64}{3}\right)H_{-1,0,0}+\left(-\frac{32}{3}\right)H_{-1,0}+\frac{64}{3}H_{0,0,0}+\frac{512}{3}H_{-2,0}+\frac{640}{3}H_3+\frac{872}{9}\right)
    +\frac{x^3}{x+1}\left(\left(-\frac{80}{3}\right)H_3+\left(-\frac{64}{3}\right)H_{-2,0}+\left(-\frac{8}{3}\right)H_{0,0,0}+\left(-\frac{4}{3}\right)H_{-1,0}+\frac{4}{3}H_{0,0}+\frac{8}{3}H_{-1,0,0}+\frac{80}{3}H_{-1,2}\right)
    +\frac{56 x^3}{3 (x+1)}\zeta _3+\left(-\frac{40 \zeta _2}{3 x}\right)H_1+\frac{40 x^2 \zeta _2}{3}H_1
    +\left(-\frac{16 \zeta _2}{x (x+1)}\right)H_{-1}\right)
 \,,
\end{dmath}

  \subsection{Time-like results in ${\overline{\text{MS}}}$}
\begin{dmath}[style={\small},compact]
\delta P_{qq}^{T(0)}=
4 C_F \left[\frac{1}{1-z}\right]_++3 C_F \delta (1-z)-4 C_F
\,,
\end{dmath}

\begin{dmath}[style={\small},compact]
\delta P_{qq}^{T(1)}=
\left[\frac{1}{1-z}\right]_+ \left(\left(\frac{268}{9}-8 \zeta _2\right) C_A C_F-\frac{80}{9} C_F N_f T_F\right)+C_A C_F \left(\frac{z}{z-1}\left(\left(-\frac{44}{3}\right)H_0-8 H_{0,0}\right)+2 z+8 \zeta _2-\frac{286}{9}\right)+C_F^2 \left(\frac{z}{z-1}\left(32 H_{0,0}+16 H_{1,0}-12 H_0+16 H_2\right)-4 z+4\right)+\delta (1-z) \left(C_A C_F \left(\frac{44}{3}\zeta _2-12 \zeta_3+\frac{17}{6}\right)+\left(\left(-\frac{16}{3}\right)\zeta _2-\frac{2}{3}\right) C_F N_f T_F+C_F^2 \left(-12 \zeta _2+24 \zeta_3+\frac{3}{2}\right)\right)+C_F N_f T_F \left(\frac{16 z}{3 (z-1)}H_0+\frac{80}{9}\right)\,,
\end{dmath}

\begin{dmath*}[style={\small},compact]
\delta P_{qq}^{T(2)}= N_f T_F C_F^2\left(\frac{64}{z-1}\zeta _3+\left(-\frac{320 z}{3 (z-1)}\right)\zeta _3+\frac{z}{z-1}\left(\left(-\frac{992}{9}\right)H_{0,0}+\left(-\frac{320}{3}\right)H_{0,0,0}+\left(-\frac{640}{9}\right)H_2+\left(-\frac{640}{9}\right)H_{1,0}+\left(-\frac{128}{3}\right)H_3+\left(-\frac{128}{3}\right)H_{2,0}+\frac{104}{3}H_0+\frac{436}{9}\right)+\frac{z^2}{z-1}\left(\frac{32}{3}H_0+\frac{112}{9}\right)+\frac{64 z \zeta _2}{3 (z-1)}H_0-\frac{548}{9 (z-1)}\right)
+C_F^3\left(\left(\left(-\frac{64}{3}\right)H_{-1,-1,0}+\left(-\frac{32}{3}\right)H_{-1,2}+\frac{8}{3}H_{-1,0}+\frac{16}{3}H_{-1,0,0}+\frac{32}{3}H_3+\frac{64}{3}H_{-2,0}\right) z^2+\left(\frac{40}{3}H_{-1,0}+\frac{80}{3}H_2+96 H_3+192 H_{-2,0}-96 H_{-1,2}+16 H_{1,0}-192 H_{-1,-1,0}+48 H_{-1,0,0}\right) z+\left(-\frac{16}{3}\right)H_2+\frac{40}{3}H_{-1,0}+\frac{1}{z-1}\left(\left(-\frac{80}{3}\right)H_0+\frac{16}{3}H_{0,0}+\frac{200}{3}H_1+\frac{104}{3}\right)+\frac{1}{z}\left(\left(-\frac{64}{3}\right)H_{-1,-1,0}+\left(-\frac{32}{3}\right)H_{-1,2}+\frac{8}{3}H_{-1,0}+\frac{16}{3}H_{-1,0,0}\right)+\left(-\frac{72 z}{z-1}\right)\zeta _4+\frac{z}{z-1}\left(\left(-\frac{400}{3}\right)H_1+\left(-\frac{146}{3}\right)H_0-448 H_4+64 H_{-3,0}-92 H_{0,0}-256 H_{1,3}+96 H_{2,0}-128 H_{2,2}-256 H_{3,0}-128 H_{3,1}-384 H_{-2,-1,0}+192 H_{-2,0,0}+336 H_{0,0,0}-256 H_{1,-2,0}+192 H_{1,0,0}-128 H_{1,2,0}-192 H_{2,0,0}-128 H_{2,1,0}-448 H_{0,0,0,0}-256 H_{1,0,0,0}-\frac{208}{3}\right)+\frac{z^2}{z-1}\left(\left(-\frac{128}{3}\right)H_{0,0,0}+\frac{112}{3}H_{0,0}+\frac{200}{3}H_1+\frac{208}{3}H_0+\frac{104}{3}\right)+\frac{128 z^2}{3 (z-1)}\zeta _3+\frac{z^3}{z-1}\left(\left(-\frac{16}{3}\right)H_{0,0,0}+\left(-\frac{8}{3}\right)H_{0,0}\right)+\frac{16 z^3}{3 (z-1)}\zeta _3+\left(-\frac{16 \zeta _2}{3 (z-1) z}\right)H_1+\frac{1}{z-1}\left(\frac{64}{3}-\frac{128 H_1}{3}\right)\zeta _2+\frac{z}{z-1}\left(32 H_0-384 H_1-48\right)\zeta _3+\frac{z}{z-1}\left(-192 H_{-2}+192 H_0+96 H_1+256 H_{0,0}-\frac{104}{3}\right)\zeta _2+\frac{z^2}{z-1}\left(-\frac{128 H_0}{3}-\frac{128 H_1}{3}+\frac{32}{3}\right)\zeta _2+\frac{z^3}{z-1}\left(-\frac{16 H_0}{3}-\frac{16 H_1}{3}+\frac{8}{3}\right)\zeta _2-96 H_{-1,2}-16 H_{1,0}-192 H_{-1,-1,0}+48 H_{-1,0,0}\right) 
\end{dmath*}
\begin{dmath*}[style={\small},compact]
+ N_f^2 T_F^2 C_F \left(-\frac{32 z}{9}+\frac{z}{z-1}\left(\left(-\frac{320}{27}\right)H_0+\left(-\frac{64}{9}\right)H_{0,0}\right)+\frac{160}{27}\right)
+C_A^2  C_F\left(4 H_{-1} \zeta _2 z^2+\left(\left(-\frac{4}{3}\right)H_{-1,0}-8 H_{-1,2}-8 H_{-1,-1,0}\right) z^2+36 H_{-1} \zeta _2 z+\left(8 H_2+4 H_{-1,0}-72 H_{-1,2}-72 H_{-1,-1,0}\right) z+\frac{1}{z-1}\left(\left(-\frac{4}{3}\right)H_0+\frac{92}{3}H_1+\frac{1586}{9}\right)+\frac{88}{3 (z-1)}\zeta _3+\frac{88}{z-1}\zeta _4+\frac{1}{z}\left(\left(-\frac{4}{3}\right)H_{-1,0}-8 H_{-1,2}-8 H_{-1,-1,0}\right)+\left(-\frac{192 z}{z-1}\right)\zeta _4+\frac{z}{z-1}\left(\left(-\frac{5396}{27}\right)H_0+\left(-\frac{1592}{9}\right)H_{0,0}+\left(-\frac{304}{3}\right)H_3+\left(-\frac{248}{3}\right)H_{0,0,0}+\left(-\frac{184}{3}\right)H_1-32 H_4-32 H_{-3,0}-24 H_{-2,0}-64 H_{-2,2}-128 H_{1,3}-128 H_{-2,-1,0}+32 H_{-2,0,0}-192 H_{1,-2,0}+88 H_{1,0,0}+64 H_{2,0,0}+32 H_{0,0,0,0}+96 H_{1,0,0,0}+128 H_{1,1,0,0}-\frac{1702}{9}\right)+\left(-\frac{32 z^2}{z-1}\right)\zeta _3+\frac{z^2}{z-1}\left(\frac{8}{3}H_{0,0}+\frac{92}{3}H_1+\frac{140}{3}H_0+64 H_3+64 H_{-2,0}+\frac{116}{9}\right)+\left(-\frac{4 z^3}{z-1}\right)\zeta _3+\frac{z^3}{z-1}\left(\frac{4}{3}H_{0,0}+8 H_3+8 H_{-2,0}\right)+\frac{4 \zeta _2}{z}H_{-1}+\left(-\frac{4 \zeta _2}{(z-1) z}\right)H_1+\frac{1}{z-1}\left(-32 H_1-\frac{1000}{9}\right)\zeta _2+\frac{z}{z-1}\left(-128 H_0-288 H_1+20\right)\zeta _3+\frac{z}{z-1}\left(\frac{464 H_0}{3}+72 H_1+32 H_{1,0}+\frac{1036}{9}\right)\zeta _2+\frac{z^2}{z-1}\left(-64 H_0-32 H_1-\frac{8}{3}\right)\zeta _2+\frac{z^3}{z-1}\left(-8 H_0-4 H_1-\frac{4}{3}\right)\zeta _2+8 H_2+36 H_{-1} \zeta _2+4 H_{-1,0}-72 H_{-1,2}-72 H_{-1,-1,0}\right)
+ N_f T_F C_A C_F\left(\left(-\frac{8}{3}\right)H_1+\left(-\frac{224}{3 (z-1)}\right)\zeta _3+\frac{320}{9 (z-1)}\zeta _2+\frac{8 z}{3}H_1+\frac{z}{z-1}\left(\frac{32}{3}H_3+\frac{64}{3}H_{0,0,0}+\frac{224}{3}H_{0,0}+\frac{2744}{27}H_0-32 H_{1,0,0}+\frac{952}{27}\right)+\frac{96 z}{z-1}\zeta _3+\frac{z^2}{z-1}\left(\left(-\frac{8}{3}\right)H_0+\frac{40}{3}\right)+\frac{z}{z-1}\left(-\frac{64 H_0}{3}-\frac{320}{9}\right)\zeta _2-\frac{1312}{27 (z-1)}\right)
+C_A  C_F^2 \bigg(\left(\left(-\frac{8}{3}\right)H_{-1,0,0}+\frac{4}{3}H_{-1,0}+\frac{64}{3}H_{-1,2}+\frac{80}{3}H_{-1,-1,0}\right) z^2-8 H_{-1} \zeta _2 z^2-72 H_{-1} \zeta _2 z+\left(\left(-\frac{44}{3}\right)H_{-1,0}+192 H_{-1,2}+240 H_{-1,-1,0}-24 H_{-1,0,0}\right) z+\left(-\frac{44}{3}\right)H_{-1,0}+\frac{1}{z-1}\left(\left(-\frac{328}{3}\right)H_1+\left(-\frac{8}{3}\right)H_{0,0}+\frac{40}{3}H_2+16 H_0-8 H_{1,0}-\frac{872}{9}\right)+\frac{1}{z}\left(\left(-\frac{8}{3}\right)H_{-1,0,0}+\frac{4}{3}H_{-1,0}+\frac{64}{3}H_{-1,2}+\frac{80}{3}H_{-1,-1,0}\right)
\end{dmath*}
\begin{dmath}[style={\small},compact]
+\frac{z}{z-1}\left(\frac{352}{3}H_{2,0}+\frac{656}{3}H_1+\frac{736}{3}H_{0,0,0}+\frac{2288}{9}H_2+\frac{2288}{9}H_{1,0}+\frac{1216}{3}H_3+\frac{3892}{9}H_{0,0}+26 H_0+160 H_4+32 H_{-3,0}+144 H_{-2,0}+128 H_{-2,2}+256 H_{1,3}+32 H_{3,0}+448 H_{-2,-1,0}-160 H_{-2,0,0}+512 H_{1,-2,0}-96 H_{1,0,0}-96 H_{2,0,0}+96 H_{0,0,0,0}-160 H_{1,0,0,0}-256 H_{1,1,0,0}+\frac{1744}{9}\right)+\frac{276 z}{z-1}\zeta _4+\frac{z^2}{z-1}\left(\left(-\frac{640}{3}\right)H_{-2,0}+\left(-\frac{512}{3}\right)H_3+\left(-\frac{428}{3}\right)H_0+\left(-\frac{328}{3}\right)H_1+\left(-\frac{88}{3}\right)H_2+\frac{64}{3}H_{0,0,0}-24 H_{0,0}-8 H_{1,0}-\frac{872}{9}\right)+\frac{128 z^2}{3 (z-1)}\zeta _3+\frac{z^3}{z-1}\left(\left(-\frac{80}{3}\right)H_{-2,0}+\left(-\frac{64}{3}\right)H_3+\left(-\frac{4}{3}\right)H_{0,0}+\frac{8}{3}H_{0,0,0}\right)+\frac{16 z^3}{3 (z-1)}\zeta _3+\left(-\frac{8 \zeta _2}{z}\right)H_{-1}+\frac{32 \zeta _2}{3 (z-1) z}H_1+\frac{1}{z-1}\left(\frac{256 H_1}{3}-\frac{80}{3}\right)\zeta _2+\frac{z}{z-1}\left(240 H_0+768 H_1-\frac{224}{3}\right)\zeta _3+\frac{z}{z-1}\left(96 H_{-2}-\frac{1040 H_0}{3}-192 H_1-64 H_2-128 H_{0,0}-64 H_{1,0}+\frac{76}{3}\right)\zeta _2+\frac{z^2}{z-1}\left(\frac{448 H_0}{3}+\frac{256 H_1}{3}\right)\zeta _2+\frac{z^3}{z-1}\left(\frac{56 H_0}{3}+\frac{32 H_1}{3}+\frac{4}{3}\right)\zeta _2-72 H_{-1} \zeta _2+192 H_{-1,2}+240 H_{-1,-1,0}-24 H_{-1,0,0}\bigg) 
+\left[\frac{1}{1-z}\right]_+ \left(C_F \left(\left(-\frac{1072}{9}\right)\zeta _2+88 \zeta _4+\frac{88 \zeta_3}{3}+\frac{490}{3}\right) C_A^2+C_F N_f T_F \left(\frac{320}{9}\zeta _2-\frac{224 \zeta_3}{3}-\frac{1672}{27}\right) C_A-\frac{64}{27} C_F N_f^2 T_F^2+C_F^2 N_f T_F \left(-\frac{220}{3}+64 \zeta_3\right)\right)
+\delta (1-z) \left(\left((18-32 \zeta_3) \zeta _2+144 \zeta _4-240 \zeta_5+68 \zeta_3+\frac{29}{2}\right) C_F^3+N_f T_F \left(\frac{40}{3}\zeta _2+\frac{232}{3}\zeta _4-\frac{272 \zeta_3}{3}-46\right) C_F^2+N_f^2 T_F^2 \left(\frac{320}{27}\zeta _2-\frac{64 \zeta_3}{9}-\frac{68}{9}\right) C_F+C_A^2 \left(\frac{4496}{27}\zeta _2-5 \zeta _4+40 \zeta_5-\frac{1552 \zeta_3}{9}-\frac{1657}{36}\right) C_F+C_A \left(\left(\left(-\frac{494}{3}\right)\zeta _4+\left(-\frac{410}{3}+16 \zeta_3\right)\zeta _2+120 \zeta_5+\frac{844 \zeta_3}{3}+\frac{151}{4}\right) C_F^2+N_f T_F \left(\left(-\frac{2672}{27}\right)\zeta _2+4 \zeta _4+\frac{400 \zeta_3}{9}+40\right) C_F\right)\right)
\,,
\end{dmath}

\begin{dmath}[style={\small},compact]
\delta P_{\bar q q}^{T(1)}=
C_A C_F \left(\frac{z}{z+1}\left(8 H_{0,0}-16 H_{-1,0}\right)-\frac{2 z^2}{z+1}+\left(-\frac{8 z}{z+1}\right)\zeta _2+\frac{2}{z+1}\right)+C_F^2 \left(\frac{z}{z+1}\left(32 H_{-1,0}-16 H_{0,0}\right)+\frac{4 z^2}{z+1}+\frac{16 z}{z+1}\zeta _2-\frac{4}{z+1}\right)
\,,
\end{dmath}

\begin{dmath*}[style={\small},compact]
\delta P_{\bar q q}^{T(2)}=
N_f T_F C_F^2\left(-\frac{112 z^2}{9 (z+1)}+\left(-\frac{32}{3}\right)H_1+\frac{32 z}{3}H_1+\frac{z}{z+1}\left(\left(-\frac{1280}{9}\right)H_{-1,0}+\left(-\frac{128}{3}\right)H_3+\left(-\frac{128}{3}\right)H_{-2,0}+\left(-\frac{128}{3}\right)H_{-1,0,0}+\frac{128}{3}H_{0,0,0}+\frac{640}{9}H_{0,0}+\frac{256}{3}H_{-1,2}\right)+\frac{64 z}{z+1}\zeta _3+\frac{z}{z+1}\left(-\frac{256 H_{-1}}{3}+\frac{64 H_0}{3}-\frac{640}{9}\right)\zeta _2+\frac{112}{9 (z+1)}\right)
+N_f T_F C_F C_A\left(\frac{56 z^2}{9 (z+1)}+\frac{16}{3}H_1+\left(-\frac{16 z}{3}\right)H_1+\left(-\frac{32 z}{z+1}\right)\zeta _3+\frac{z}{z+1}\left(\left(-\frac{128}{3}\right)H_{-1,2}+\left(-\frac{320}{9}\right)H_{0,0}+\left(-\frac{64}{3}\right)H_{0,0,0}+\frac{64}{3}H_3+\frac{64}{3}H_{-2,0}+\frac{64}{3}H_{-1,0,0}+\frac{640}{9}H_{-1,0}\right)+\frac{z}{z+1}\left(\frac{128 H_{-1}}{3}-\frac{32 H_0}{3}+\frac{320}{9}\right)\zeta _2-\frac{56}{9 (z+1)}\right) 
\end{dmath*}
\begin{dmath*}[style={\small},compact]
+
C_F C_A^2 \left(-4 H_1 \zeta _2 z^2-8 H_{-1,-1,0} z^2+36 H_1 \zeta _2 z+\left(\frac{68}{3}H_1+72 H_{-1,-1,0}\right) z+\left(-\frac{68}{3}\right)H_1+\left(-\frac{8}{z}\right)H_{-1,-1,0}+\frac{1}{z+1}\left(\left(-\frac{4}{3}\right)H_0+\frac{16}{3}H_{-1,0}+8 H_2+64 H_{-1,2}+\frac{358}{9}\right)+\frac{1}{z (z+1)}\left(\frac{4}{3}H_{-1,0}-8 H_{-1,2}\right)+\frac{z}{z+1}\left(\left(-\frac{2072}{9}\right)H_{-1,0}+\left(-\frac{392}{3}\right)H_3+\left(-\frac{392}{3}\right)H_{-2,0}+\left(-\frac{176}{3}\right)H_{-1,0,0}+\frac{88}{3}H_0+\frac{248}{3}H_{0,0,0}+\frac{1108}{9}H_{0,0}+\frac{784}{3}H_{-1,2}+16 H_2-64 H_4+32 H_{-3,0}+256 H_{-2,2}+256 H_{-1,3}+128 H_{-2,0,0}-512 H_{-1,-1,2}-256 H_{-1,-1,0,0}+96 H_{-1,0,0,0}-32 H_{0,0,0,0}\right)+\frac{80 z}{z+1}\zeta _4+\frac{z^2}{z+1}\left(\frac{8}{3}H_{0,0}+\frac{16}{3}H_{-1,0}+\frac{20}{3}H_0+8 H_2-64 H_3-64 H_{-2,0}+64 H_{-1,2}-\frac{358}{9}\right)+\frac{32 z^2}{z+1}\zeta _3+\left(-\frac{4 z^3}{z+1}\right)\zeta _3+\frac{z^3}{z+1}\left(\left(-\frac{4}{3}\right)H_{0,0}+\frac{4}{3}H_{-1,0}+8 H_3+8 H_{-2,0}-8 H_{-1,2}\right)+\frac{4 \zeta _2}{z}H_1+\frac{4 \zeta _2}{z (z+1)}H_{-1}+\frac{1}{z+1}\left(-32 H_{-1}-8\right)\zeta _2+\frac{z}{z+1}\left(-384 H_{-1}+128 H_0+124\right)\zeta _3+\frac{z}{z+1}\left(-256 H_{-2}-\frac{568 H_{-1}}{3}+\frac{232 H_0}{3}-64 H_2+512 H_{-1,-1}-352 H_{-1,0}+96 H_{0,0}-\frac{1180}{9}\right)\zeta _2+\frac{z^2}{z+1}\left(-32 H_{-1}+64 H_0-\frac{8}{3}\right)\zeta _2+\frac{z^3}{z+1}\left(4 H_{-1}-8 H_0+\frac{4}{3}\right)\zeta _2-36 H_1 \zeta _2+72 H_{-1,-1,0}\right) 
+C_F^3 \bigg(\left(-\frac{184}{3}\right)H_1+\left(-\frac{32}{3 z}\right)H_{-1,-1,0}+\left(-\frac{32 z^2}{3}\right)H_{-1,-1,0}+\frac{1}{z+1}\left(\left(-\frac{80}{3}\right)H_0+\frac{16}{3}H_{0,0}+\frac{112}{3}H_2+\frac{128}{3}H_{-1,0,0}+\frac{512}{3}H_{-1,2}+\frac{104}{3}\right)+\frac{1}{z (z+1)}\left(\left(-\frac{64}{3}\right)H_{-1,2}+\left(-\frac{16}{3}\right)H_{-1,0,0}+\left(-\frac{8}{3}\right)H_{-1,0}\right)+\left(-\frac{280 z}{z+1}\right)\zeta _4
+\frac{z}{z+1}\left(\left(-\frac{40}{3}\right)H_{0,0}+\frac{16}{3}H_{-1,0}+\frac{128}{3}H_2+80 H_0-288 H_3-320 H_{-3,0}+640 H_{-2,2}+576 H_{-1,2}+512 H_{-1,3}+64 H_{3,0}+256 H_{-2,-1,0}-64 H_{-2,0,0}+256 H_{-1,-2,0}-1536 H_{-1,-1,2}+384 H_{-1,0,0}-128 H_{-1,2,0}-192 H_{0,0,0}-256 H_{-1,-1,0,0}-448 H_{-1,0,0,0}+320 H_{0,0,0,0}\right)+\frac{z^2}{z+1}\left(\left(-\frac{512}{3}\right)H_3+\left(-\frac{256}{3}\right)H_{-2,0}+\left(-\frac{128}{3}\right)H_{0,0,0}+\frac{16}{3}H_2+\frac{128}{3}H_{-1,0,0}+\frac{176}{3}H_0+\frac{512}{3}H_{-1,2}-16 H_{0,0}-\frac{104}{3}\right)
+\frac{512 z^2}{3 (z+1)}\zeta _3+\left(-\frac{64 z^3}{3 (z+1)}\right)\zeta _3
+\frac{z^3}{z+1}\left(\left(-\frac{64}{3}\right)H_{-1,2}+\left(-\frac{16}{3}\right)H_{-1,0,0}+\left(-\frac{8}{3}\right)H_{-1,0}+\frac{8}{3}H_{0,0}+\frac{16}{3}H_{0,0,0}+\frac{32}{3}H_{-2,0}+\frac{64}{3}H_3\right)+\frac{32 \zeta _2}{3 z}H_1+\left(-\frac{32}{3} z^2 \zeta _2\right)H_1+\frac{16 \zeta _2}{z (z+1)}H_{-1}+\frac{1}{z+1}\left(-128 H_{-1}-\frac{32}{3}\right)\zeta _2+\frac{z}{z+1}\left(-1152 H_{-1}-32 H_0+336\right)\zeta _3+\frac{z}{z+1}\left(-512 H_{-2}-480 H_{-1}+384 H_0-192 H_2+1536 H_{-1,-1}-768 H_{-1,0}-64 H_{0,0}-40\right)\zeta _2+\frac{z^2}{z+1}\left(-128 H_{-1}+\frac{640 H_0}{3}-32\right)\zeta _2+\frac{z^3}{z+1}\left(16 H_{-1}-\frac{80 H_0}{3}-\frac{8}{3}\right)\zeta _2+96 z H_1 \zeta _2-96 H_1 \zeta _2+16 H_{1,0}+96 H_{-1,-1,0}+z \left(\frac{184}{3}H_1-16 H_{1,0}+96 H_{-1,-1,0}\right)\bigg) 
\end{dmath*}
\begin{dmath}[style={\small},compact]
+  C_A C_F^2\left(\frac{64}{3 z}H_{-1,-1,0}+\frac{64 z^2}{3}H_{-1,-1,0}+\frac{1}{z+1}\left(\left(-\frac{640}{3}\right)H_{-1,2}+\left(-\frac{104}{3}\right)H_2+\left(-\frac{64}{3}\right)H_{-1,0,0}+\left(-\frac{32}{3}\right)H_{-1,0}+\left(-\frac{8}{3}\right)H_{0,0}+16 H_0-\frac{872}{9}\right)+\frac{1}{z (z+1)}\left(\left(-\frac{4}{3}\right)H_{-1,0}+\frac{8}{3}H_{-1,0,0}+\frac{80}{3}H_{-1,2}\right)+\left(-\frac{20 z}{z+1}\right)\zeta _4+\frac{z}{z+1}\left(\left(-\frac{2432}{3}\right)H_{-1,2}+\left(-\frac{2156}{9}\right)H_{0,0}+\left(-\frac{296}{3}\right)H_0+\left(-\frac{224}{3}\right)H_{-1,0,0}+\left(-\frac{208}{3}\right)H_{0,0,0}+\left(-\frac{160}{3}\right)H_2+\frac{784}{3}H_{-2,0}+\frac{1216}{3}H_3+\frac{4120}{9}H_{-1,0}+128 H_4+96 H_{-3,0}-832 H_{-2,2}-768 H_{-1,3}-32 H_{3,0}-128 H_{-2,-1,0}-224 H_{-2,0,0}-128 H_{-1,-2,0}+1792 H_{-1,-1,2}+64 H_{-1,2,0}+640 H_{-1,-1,0,0}+32 H_{-1,0,0,0}-96 H_{0,0,0,0}\right)+\left(-\frac{448 z^2}{3 (z+1)}\right)\zeta _3+\frac{z^2}{z+1}\left(\left(-\frac{640}{3}\right)H_{-1,2}+\left(-\frac{128}{3}\right)H_0+\left(-\frac{64}{3}\right)H_{-1,0,0}+\left(-\frac{56}{3}\right)H_2+\left(-\frac{32}{3}\right)H_{-1,0}+\frac{8}{3}H_{0,0}+\frac{64}{3}H_{0,0,0}+\frac{512}{3}H_{-2,0}+\frac{640}{3}H_3+\frac{872}{9}\right)+\frac{z^3}{z+1}\left(\left(-\frac{80}{3}\right)H_3+\left(-\frac{64}{3}\right)H_{-2,0}+\left(-\frac{8}{3}\right)H_{0,0,0}+\left(-\frac{4}{3}\right)H_{-1,0}+\frac{4}{3}H_{0,0}+\frac{8}{3}H_{-1,0,0}+\frac{80}{3}H_{-1,2}\right)+\frac{56 z^3}{3 (z+1)}\zeta _3+\left(-\frac{40 \zeta _2}{3 z}\right)H_1+\frac{40 z^2 \zeta _2}{3}H_1+\left(-\frac{16 \zeta _2}{z (z+1)}\right)H_{-1}+\frac{1}{z+1}\left(128 H_{-1}+\frac{64}{3}\right)\zeta _2+\frac{z}{z+1}\left(1344 H_{-1}-240 H_0-416\right)\zeta _3+\frac{z}{z+1}\left(768 H_{-2}+\frac{1856 H_{-1}}{3}-\frac{1040 H_0}{3}+224 H_2-1792 H_{-1,-1}+1088 H_{-1,0}-160 H_{0,0}+\frac{2540}{9}\right)\zeta _2+\frac{z^2}{z+1}\left(128 H_{-1}-\frac{704 H_0}{3}+\frac{64}{3}\right)\zeta _2+\frac{z^3}{z+1}\left(-16 H_{-1}+\frac{88 H_0}{3}-\frac{4}{3}\right)\zeta _2+76 H_1-120 z H_1 \zeta _2+120 H_1 \zeta _2-8 H_{1,0}+z \left(-76 H_1+8 H_{1,0}-192 H_{-1,-1,0}\right)-192 H_{-1,-1,0}\right)\,.
\end{dmath}

\section{N$^3$LO coefficients for  transversity  TMDs in ${\overline{\text{MS}}}$}
\label{sec:n3lo-coeff-transversity}
In this section, we present the  N$^3$LO matching coefficients for transversity TMDs in the ${\overline{\text{MS}}}$ scheme. 
Particular attention is paid to their behavior in the end-point regions, where logarithmically enhanced terms arise and affect the convergence of the perturbative expansion. 
These structures highlight the importance of end-point resummation for achieving improved perturbative stability and reliable phenomenological applications. 
A detailed study of such resummation effects is left for future investigations.
 In the following sections, we collect the leading-power Regge and threshold limits of the coefficient functions, extracted from our analytic results provided in the ancillary files accompanying the arXiv submission.

\subsection{Regge and threshold limits}
\label{sec:endpoint}
The coefficient functions develop end-point divergences both in the threshold and high energy limits.
We first present here the results for leading threshold limit. The results for high energy limit will be discussed in next section.
In the $z \to 1$ limit, we have exactly the same results as the unpolarized case~\cite{Luo:2019szz,Luo:2020epw}
\begin{equation}
  \label{eq:largez}
 \lim_{z\to1} \delta \mathcal{I}^{(2)}_{ij}(z)= \lim_{z\to1}\delta \mathcal{C}^{(2)}_{ji}(z) = \frac{2 \gamma^R_{1,i}}{(1-z)_+}\delta_{ij} \,, \quad 
  \lim_{z\to1} \delta \mathcal{I}^{(3)}_{ij}(z)=\lim_{z\to1}\delta \mathcal{C}^{(3)}_{ji}(z) = \frac{2 \gamma^R_{2,i}}{(1-z)_+} \delta_{ij} \,,
\end{equation}
where $\gamma^R_{1(2)}$ are the two(three)-loop rapidity anomalous dimensions~\cite{Li:2016ctv,Vladimirov:2016dll}. The relation between threshold limit and rapidity anomalous dimension has been anticipated in \cite{Echevarria:2016scs,Lustermans:2016nvk,Billis:2019vxg}.
The explicit expressions up to three-loop read~\cite{Luo:2019szz}
 \begin{align}
\delta\mathcal{I}^{(1)}_{qq}(z)\overset{z\to 1}{=}&\delta\mathcal{C}^{(1)}_{qq}(z)\simeq 0\,,
 \nn\\
\delta\mathcal{I}^{(2)}_{qq}(z)\overset{z\to 1}{=}&\delta\mathcal{C}^{(2)}_{qq}(z)\simeq  \frac{1}{(1-z)_+} \bigg[ \left(28 \zeta _3-\frac{808}{27}\right) C_A C_F+\frac{224}{27} C_F N_f T_F \bigg] \,,
\nn \\
\delta\mathcal{I}^{(3)}_{qq}(z)\overset{z\to 1}{=}&\delta\mathcal{C}^{(3)}_{qq}(z)\simeq  \frac{1}{(1-z)_+} \bigg[ 
   \biggl(-\frac{1648 \zeta _2}{81}-\frac{1808 \zeta _3}{27}+\frac{40 \zeta
   _4}{3} 
   +\frac{125252}{729}\biggl) C_A C_F N_f T_F\nonumber 
    \\
   +&\biggl(-\frac{176}{3} \zeta _3 \zeta
   _2+\frac{6392 \zeta _2}{81}  
 +\frac{12328 \zeta _3}{27}+\frac{154 \zeta _4}{3}-192 \zeta
   _5-\frac{297029}{729}\biggl) C_A^2 C_F \nonumber 
   \\
   +& \left(-\frac{608 \zeta _3}{9}-32 \zeta
   _4+\frac{3422}{27}\right) C_F^2 N_f T_F
  + \left(-\frac{128}{9} \zeta
   _3-\frac{7424}{729}\right) C_F N_f^2 T_F^2  
     \bigg] \,.
 \end{align}
It is also instructive to investigate the small-$x$ behavior of the transversity coefficient functions and to compare them with the corresponding unpolarized case. Unlike the unpolarized coefficient functions, which exhibit a power-like divergence $\sim 1/x$ in the small-$x$ limit, the transversity coefficient functions diverge only logarithmically. For clarity, we summarize below the leading power small-$x$ limit of the coefficient functions. In particular, we focus on the scale-independent functions $\delta I^{(n)}_{ji} (x)$ and  $\delta C^{(n)}_{ji} (x)$ that appear in the RG solutions, cf. Eqs.~(\ref{eq:RGs-1}) and (\ref{eq:RGs-2}).
\subsubsection{Small-$x$ limit of  the coefficient functions for TMD transversity PDFs}
\label{sec:small-x-expansion-PDF}
\begin{align}
\delta I^{ (1)}_{qq} (x)\simeq&
0
\,,
\end{align}

\begin{align}
\delta I^{ (2)}_{qq} (x)\simeq&
\frac{14 C_A C_F}{3}-\frac{4}{3} C_F N_f T_F-2 C_F^2
\,,
\end{align}

\begin{align}
\delta I^{ (3)}_{qq} (x)\simeq&
C_A C_F N_f T_F \left(\left(-\frac{8}{9}\right)\ln ^2x+\left(-\frac{200}{27}\right)\ln x+\frac{16}{9}\zeta _2-\frac{2672}{81}\right)
\nn\\
+&C_A C_F^2 \left(\left(-\frac{16}{3}\right)\ln ^2x+\left(-\frac{148}{9}\right)\ln x-16 \zeta _2+24 \zeta _3-\frac{12616}{81}\right)
\nn\\
+&C_A^2 C_F \left(\frac{22}{9}\ln ^2x+\frac{296}{27}\ln x+\left(-\frac{44}{3}\right)\zeta _3+\frac{4}{9}\zeta _2+\frac{2384}{27}\right)
\nn\\
+&C_F^2 N_f T_F \left(\frac{16}{9}\ln ^2x+\frac{64}{9}\ln x+\frac{2600}{81}\right)+C_F N_f^2 T_F^2 \left(\frac{32}{27}\ln x+\frac{160}{81}\right)
\nn\\
+&C_F^3 \left(\frac{8}{9}\ln ^2x+\left(-\frac{40}{9}\right)\ln x+\frac{32}{3}\zeta _2+\frac{32}{3}\zeta _3+\frac{464}{9}\right)
\,,
\end{align}

\begin{align}
\delta I^{ (1)}_{q\bar q} (x)\simeq&
0
\,,
\end{align}

\begin{align}
\delta I^{ (2)}_{q\bar q} (x)\simeq&
2 C_F^2-C_A C_F
\,,
\end{align}

\begin{align}
\delta I^{ (3)}_{q\bar q} (x)\simeq&
 C_A C_F N_f T_F \left(\frac{8}{9}\ln ^2x+\frac{8}{9}\ln x+\left(-\frac{16}{9}\right)\zeta _2+\frac{1192}{81}\right)
 \nn\\
 +&C_A C_F^2 \left(\frac{16}{3}\ln ^2x+\frac{16}{9}\ln x+\left(-\frac{232}{9}\right)\zeta _2-24 \zeta _3+\frac{11428}{81}\right)
 \nn\\
 +&C_A^2 C_F \left(\left(-\frac{22}{9}\right)\ln ^2x-2 \ln x+\frac{92}{9}\zeta _2+\frac{44}{3}\zeta _3-\frac{4670}{81}\right)
 \nn\\
+&C_F^2 N_f T_F \left(\left(-\frac{16}{9}\right)\ln ^2x+\left(-\frac{16}{9}\right)\ln x+\frac{32}{9}\zeta _2-\frac{2384}{81}\right)
\nn\\
+&C_F^3 \left(\left(-\frac{8}{9}\right)\ln ^2x+\frac{40}{9}\ln x+\left(-\frac{32}{3}\right)\zeta _3+\frac{32}{3}\zeta _2-\frac{464}{9}\right)
\,.
\end{align}

\subsubsection{Small-$z$ limit of the coefficient functions for TMD transversity  FFs}
\label{sec:small-x-expansion-FF}
\begin{align}
\delta C^{ (1)}_{qq} (z)\simeq&
0
\,,
\end{align}

\begin{align}
\delta C^{ (2)}_{qq} (z)\simeq&
C_A C_F \left(\frac{14}{3}-2 \ln z\right)-\frac{4}{3} C_F N_f T_F+C_F^2 (4 \ln z-2)
\,,
\end{align}

\begin{align}
\delta C^{ (3)}_{qq} (z)\simeq&
C_A C_F N_f T_F \left(\left(-\frac{32}{9}\right)\ln ^2z+\frac{40}{27}\ln z+\frac{16}{9}\zeta _2-\frac{2672}{81}\right)
\nn\\
+&C_A C_F^2 \left(\left(\left(-\frac{56}{3}\right)\zeta _2+\frac{310}{3}\right) \ln z+\left(-\frac{260}{9}\right)\ln ^2z-116 \zeta _2-\frac{11932}{81}\right)
\nn\\
+&C_A^2 C_F \left(\left(8 \zeta _2-\frac{838}{27}\right) \ln z+\frac{88}{9}\ln ^2z+\frac{172}{9}\zeta _2+\frac{2348}{27}\right)
\nn\\
+&C_F^2 N_f T_F \left(\frac{64}{9}\ln ^2z-16 \ln z+\frac{32}{3}\zeta _2+\frac{2600}{81}\right)
+C_F N_f^2 T_F^2 \left(\frac{32}{27}\ln z+\frac{160}{81}\right)
\nn\\
+&C_F^3 \left(\left(\frac{16}{3}\zeta _2-\frac{140}{3}\right) \ln z+\frac{56}{3}\ln ^2z+\frac{232}{3}\zeta _2+40\right)
\,,
\end{align}

\begin{align}
\delta C^{ (1)}_{\bar q q} (z)\simeq&
0
\,,
\end{align}

\begin{align}
\delta C^{ (2)}_{\bar q q} (z)\simeq&
C_A C_F (2 \ln z-1)+C_F^2 (2-4 \ln z)
\,,
\end{align}

\begin{align}
\delta C^{ (3)}_{\bar q q} (z)\simeq&
C_A C_F N_f T_F \left(\frac{32}{9}\ln ^2z-8 \ln z+\left(-\frac{64}{9}\right)\zeta _2+\frac{1192}{81}\right)
\nn\\
+&C_A C_F^2 \left(\left(\left(-\frac{88}{3}\right)\zeta _2-\frac{310}{3}\right) \ln z+\frac{260}{9}\ln ^2z+\frac{32}{9}\zeta _2+\frac{10744}{81}\right)
\nn\\
+&C_A^2 C_F \left(\left(8 \zeta _2+40\right) \ln z+\left(-\frac{88}{9}\right)\ln ^2z+\frac{56}{9}\zeta _2-\frac{4562}{81}\right)
\nn\\
+&C_F^2 N_f T_F \left(\left(-\frac{64}{9}\right)\ln ^2z+16 \ln z+\frac{128}{9}\zeta _2-\frac{2384}{81}\right)
\nn\\
+&C_F^3 \left(\left(\frac{80}{3}\zeta _2+\frac{140}{3}\right) \ln z+\left(-\frac{56}{3}\right)\ln ^2z-32 \zeta _2-40\right)
\,.
\end{align}

\subsection{Numerical fits of the N{}$^3$LO coefficients}
\label{sec:num-fit}
The NLO and NNLO space-like coefficient functions were obtained in Refs.~\cite{Buffing:2017mqm,Bacchetta:2013pqa,Gutierrez-Reyes:2017glx,Gutierrez-Reyes:2018iod}. 
The present work is devoted to extending these results to higher orders, and for the first time we provide the complete expressions up to $\text{N}^3$LO. 
By comparing our results for the regulator-independent TMD PDFs and TMD FFs with those reported in Ref.~\cite{Gutierrez-Reyes:2018iod}, 
we confirm agreement in the $\bar q \to q$ (and $q \to \bar q$) channel. For the $q \to q$ channel, however, we observe minor discrepancies 
\begin{align}
\delta \mathscr{I}^{(2)}_{qq}|-\delta \mathscr{I}^{(2)}_{qq}|_{\text{DIA}}=&-6 C_F^2(1-x)\,,
\nn\\
\delta \mathscr{C}^{(2)}_{qq}|-\delta \mathscr{C}^{(2)}_{qq}|_{\text{DIA}}=&-6 C_F^2(1-z)-\frac{2}{3}\pi^4 C_A C_F \delta(1-z)\,.
\end{align}
The analytic expressions of $\text{N}^{3}$LO coefficient functions contain harmonic polylogarithms (HPLs) up to transcendental weight $5$.
Their numerical evaluation is performed using the \texttt{HPL} package~\cite{Maitre:2005uu}.
To facilitate straightforward numerical implementation, we carry out numerical fits for all the coefficient functions, as detailed below.
Following Ref.~\cite{Moch:2017uml}, we use the following elementary functions to fit the results,
\begin{align}
\label{eq:Lzdefinition}
L_x \equiv \ln x\,,\,L_{\bar{x}} \equiv \ln (1-x )\,, \, \bar{x}\equiv 1-x \,. 
\end{align}
For NNLO and $\text{N}^{3}$LO coefficient functions, we fit the exact results in the region $10^{-6} < x <1-10^{-6}$, and we have set the color factor to their numerical values in QCD, i.e. 
\begin{align}
C_F = \frac{4}{3}\,,\qquad  C_A = 3\,, \qquad T_F = \frac{1}{2} \,. 
\end{align}
In more detail, we subtract the $x \to 0$ and $x \to 1$ limits and then fit the remaining terms in the region $10^{-6} < x <1-10^{-6}$.
 Combining the two parts, the fitted results can achieve an accuracy better than $10^{-3}$ over the range $0<x<1$. 
 Below, we provide numerical fits for the scale-independent coefficient functions 
$\delta I^{(n)}_{ji} (x)$ and  $\delta C^{(n)}_{ji} (x)$, which enter the RG solutions given in Eqs.~(\ref{eq:RGs-1}) and~(\ref{eq:RGs-2}), respectively.
 The complete numerical fits and analytic expressions  are provided in the ancillary files accompanying the arXiv submission.

\subsubsection{Numerical fit for TMD PDFs}
\label{sec:num-I}
\begin{dmath}[style={\small},compact, indentstep={0 em}]
\delta I^{(1)}_{qq}(x) =
0
\,,
\end{dmath}

\begin{dmath}[style={\small},compact, indentstep={0 em}]
\delta I^{(2)}_{qq}(x) =
N_f \left(5.53086 \left[\frac{1}{\bar{x}}\right]_++x^3 \left(0.159353 L_x^2+1.83868 L_x-3.46249\right)+x^2 \left(0.918768 L_x^2+3.30005 L_x+1.07696\right)+x \left(0.888889 L_x^2+2.96296 L_x+2.37017\right)-0.0100775 x^6+0.090776 x^5-0.657935 x^4+1.48148 \bar{x}-7.90123\right)+14.9267 \left[\frac{1}{\bar{x}}\right]_++x^3 \left(2.43688 L_x^3-17.6078 L_x^2+43.7086 L_x-39.0133\right)+x^2 \left(-2.54756 L_x^3-13.8299 L_x^2-7.50098 L_x+63.2888\right)+x \left(-2.66667 L_x^3-9.33333 L_x^2-39.1111 L_x-16.4544\right)+\bar{x}^3 \left(0.508876 L_{\bar{x}}^2-0.601064 L_{\bar{x}}\right)+\bar{x}^2 \left(1.19128 L_{\bar{x}}^2-2.40075 L_{\bar{x}}\right)-7.11111 L_{\bar{x}}^2+22.2222 L_{\bar{x}}+\bar{x} \left(3.55556 L_{\bar{x}}^2-13.3333 L_{\bar{x}}+10.0354\right)+0.0762313 x^6-0.683931 x^5+3.19561 x^4-9.851
\,,
\end{dmath}

\begin{dmath}[style={\small},compact, indentstep={0 em}]
\delta I^{(3)}_{qq}(x) =
N_f \left(154.257 \left[\frac{1}{\bar{x}}\right]_++x^3 \left(-10.139 L_x^4+53.293 L_x^3-362.551 L_x^2+1069.36 L_x-2080.42\right)+x^2 \left(3.35283 L_x^4+38.0541 L_x^3+204.028 L_x^2+893.7 L_x+1807.17\right)+x \left(3.16049 L_x^4+27.0398 L_x^3+95.569 L_x^2+243.104 L_x+167.326\right)-0.197531 L_x^2-8.49383 L_x+\bar{x}^3 \left(0.0903308 L_{\bar{x}}^3-0.534541 L_{\bar{x}}^2+1.48298 L_{\bar{x}}\right)+\bar{x}^2 \left(-0.344655 L_{\bar{x}}^3-1.88691 L_{\bar{x}}^2-4.7445 L_{\bar{x}}\right)+2.107 L_{\bar{x}}^3+10.3704 L_{\bar{x}}^2-10.4458 L_{\bar{x}}+\bar{x} \left(-1.0535 L_{\bar{x}}^3-11.358 L_{\bar{x}}^2+3.08951 L_{\bar{x}}+132.\right)-0.0981767 x^6-0.236655 x^5+65.9262 x^4-317.852\right)+N_f^2 \left(-9.09324 \left[\frac{1}{\bar{x}}\right]_++x^3 \left(0.431032 L_x^3-3.50262 L_x^2+9.32163 L_x-8.91467\right)+x^2 \left(-0.62563 L_x^3-2.59739 L_x^2-1.44432 L_x+15.2779\right)+x \left(-0.658436 L_x^3-3.29218 L_x^2-7.2428 L_x-6.71585\right)+0.395062 L_x+0.00672043 x^6-0.0760913 x^5+0.55363 x^4-6.05761 \bar{x}+15.8093\right)+140.136 \left[\frac{1}{\bar{x}}\right]_++x^3 \left(32.8538 L_x^5-59.8157 L_x^4+1263.35 L_x^3-1593.34 L_x^2+9101.99 L_x+965.588\right)+x^2 \left(-3.03843 L_x^5-67.7124 L_x^4-364.242 L_x^3-1888.44 L_x^2-2473.02 L_x+836.39\right)+x \left(-2.64691 L_x^5-37.1111 L_x^4-137.094 L_x^3-453.207 L_x^2-173.141 L_x+1583.11\right)+3.25926 L_x^2+32.5926 L_x+\bar{x}^3 \left(20.7632 L_{\bar{x}}^3-16.7047 L_{\bar{x}}^2+301.823 L_{\bar{x}}\right)+\bar{x}^2 \left(6.41987 L_{\bar{x}}^3+1.21958 L_{\bar{x}}^2+202.422 L_{\bar{x}}\right)-34.7654 L_{\bar{x}}^3-5.09037 L_{\bar{x}}^2+637.843 L_{\bar{x}}+\bar{x} \left(-7.90123 L_{\bar{x}}^3+110.483 L_{\bar{x}}^2-62.9396 L_{\bar{x}}-863.968\right)-8.32785 x^6+84.9876 x^5-3254.01 x^4+948.568
\,,
\end{dmath}

\begin{dmath}[style={\small},compact, indentstep={0 em}]
\delta I^{(2)}_{q\bar q}(x) =
x \left(2.14474\, -0.296296 L_x^3\right)+x^3 \left(15.5499 L_x^3-48.8464 L_x^2+242.942 L_x-320.161\right)+x^2 \left(1.20874 L_x^3+19.8823 L_x^2+132.773 L_x+344.416\right)+0.243961 x^6-1.09739 x^5-25.5468 x^4-0.444444 \bar{x}
\,,
\end{dmath}

\begin{dmath}[style={\small},compact, indentstep={0 em}]
\delta I^{(3)}_{q\bar q}(x) =
N_f \left(x^3 \left(-86.9687 L_x^4+339.693 L_x^3-3577.27 L_x^2+9245.24 L_x-18136.9\right)+x^2 \left(1.88363 L_x^4+69.5867 L_x^3+974.23 L_x^2+6391.29 L_x+16891.9\right)+x \left(0.263374 L_x^4+1.09739 L_x^3+5.25509 L_x^2+2.52335 L_x-14.4093\right)+0.197531 L_x^2+0.197531 L_x+\bar{x}^3 \left(0.672455 L_{\bar{x}}^2+0.57305 L_{\bar{x}}\right)+\bar{x}^2 \left(-0.0270463 L_{\bar{x}}^2-1.49294 L_{\bar{x}}\right)+\bar{x} \left(0.197531 L_{\bar{x}}^2+1.0535 L_{\bar{x}}+2.87517\right)+1.23754 x^6-39.0649 x^5+1297.47 x^4-0.254789\right)+x^3 \left(218.515 L_x^5+99.2151 L_x^4+9703.19 L_x^3-1433.99 L_x^2+66557.2 L_x+18184.5\right)+x^2 \left(-1.98473 L_x^5-69.4832 L_x^4-1029.42 L_x^3-6944.15 L_x^2-15843. L_x+16446.3\right)+x \left(-0.118519 L_x^5-3.53086 L_x^4-22.2993 L_x^3-32.9084 L_x^2+32.298 L_x+174.427\right)-3.25926 L_x^2-3.25926 L_x+\bar{x}^3 \left(1.23696 L_{\bar{x}}^2+9.1761 L_{\bar{x}}\right)+\bar{x}^2 \left(13.7472 L_{\bar{x}}-0.884454 L_{\bar{x}}^2\right)+\bar{x} \left(-3.13186 L_{\bar{x}}^2-19.7773 L_{\bar{x}}-47.6421\right)-56.825 x^6+1077.22 x^5-35852.2 x^4+26.6075
\,.
\end{dmath}

\subsubsection{Numerical fit for TMD FFs}
\label{sec:num-C}
\begin{dmath}[style={\small},compact]
\delta C^{(1)}_{qq}(z) =
z^3 \left(6.88266 L_z-7.0639\right)+z^2 \left(10.467 L_z-0.842818\right)+z \left(10.6667 L_z+5.3312\right)-0.128696 z^6+0.818297 z^5-3.44742 z^4+5.33333 \bar{z}-5.33333
\,,
\end{dmath}

\begin{dmath}[style={\small},compact]
\delta C^{(2)}_{qq}(z) =
N_f \left(5.53086 \left[\frac{1}{\bar{z}}\right]_++z^3 \left(1.70397 L_z^2-4.64799 L_z+10.359\right)+z^2 \left(0.881779 L_z^2-8.99257 L_z-0.401235\right)+z \left(0.888889 L_z^2-8.88889 L_z-3.55556\right)-0.00933333 z^6+0.104686 z^5-1.16427 z^4-4.44444 \bar{z}-1.97531\right)+14.9267 \left[\frac{1}{\bar{z}}\right]_++z^3 \left(-9.54698 L_z^3-53.3203 L_z^2+15.1648 L_z-156.471\right)+z^2 \left(-22.9556 L_z^3-10.3136 L_z^2-68.1265 L_z+154.308\right)+z \left(-23.4074 L_z^3+12. L_z^2-17.7527 L_z-99.8513\right)-0.888889 L_z+\bar{z}^3 \left(-0.799338 L_{\bar{z}}^2-1.59409 L_{\bar{z}}\right)+\bar{z}^2 \left(2.65626 L_{\bar{z}}-1.16785 L_{\bar{z}}^2\right)+7.11111 L_{\bar{z}}^2-22.2222 L_{\bar{z}}+\bar{z} \left(-3.55556 L_{\bar{z}}^2+51.5556 L_{\bar{z}}+149.849\right)+0.905556 z^6-8.00011 z^5+91.7047 z^4-149.664
\,,
\end{dmath}

\begin{dmath*}[style={\small},compact]
\delta C^{(3)}_{qq}(z) =
N_f \left(154.257 \left[\frac{1}{\bar{z}}\right]_++z^3 \left(-19.7868 L_z^4+33.0601 L_z^3-442.868 L_z^2+663.837 L_z-1896.52\right)+z^2 \left(-1.97195 L_z^4+32.7379 L_z^3+194.866 L_z^2+441.17 L_z+1895.42\right)+z \left(-2.13992 L_z^4+17.8217 L_z^3+86.4094 L_z^2-297.674 L_z-60.271\right)-0.790123 L_z^2-11.2593 L_z+\bar{z}^3 \left(-0.141494 L_{\bar{z}}^3+2.07513 L_{\bar{z}}^2-9.48914 L_{\bar{z}}\right)+\bar{z}^2 \left(0.342792 L_{\bar{z}}^3+1.84641 L_{\bar{z}}^2-25.477 L_{\bar{z}}\right)-2.107 L_{\bar{z}}^3-10.3704 L_{\bar{z}}^2+13.8356 L_{\bar{z}}+\bar{z} \left(1.0535 L_{\bar{z}}^3-7.30864 L_{\bar{z}}^2-71.9318 L_{\bar{z}}-212.901\right)+0.577185 z^6-9.72231 z^5+302.016 z^4+42.6462\right)
+N_f^2 \left(-9.09324 \left[\frac{1}{\bar{z}}\right]_++z^3 \left(1.13739 L_z^3-1.11292 L_z^2+8.49447 L_z+0.132844\right)+z^2 \left(0.911883 L_z^3-3.47057 L_z^2+5.15991 L_z-2.6942\right)+z \left(0.921811 L_z^3-3.29218 L_z^2+5.92593 L_z-0.131794\right)+0.395062 L_z-0.0139459 z^6+0.212281 z^5-3.95786 z^4+0.526749 \bar{z}+9.22493\right)
\end{dmath*}
\begin{dmath}[style={\small},compact]
+140.136 \left[\frac{1}{\bar{z}}\right]_++z^3 \left(47.894 L_z^5+302.009 L_z^4+1727.78 L_z^3+2836.22 L_z^2+18851.7 L_z-138.226\right)+z^2 \left(13.272 L_z^5+26.3528 L_z^4-155.496 L_z^3-1693.73 L_z^2+10289.1 L_z-417.803\right)+z \left(13.3136 L_z^5-22.963 L_z^4-302.296 L_z^3+602.58 L_z^2+5948.2 L_z+9896.07\right)+7.11111 L_z^2+84.0824 L_z+\bar{z}^3 \left(29.1314 L_{\bar{z}}^3-45.6048 L_{\bar{z}}^2+386.789 L_{\bar{z}}\right)+\bar{z}^2 \left(-4.9028 L_{\bar{z}}^3-96.5395 L_{\bar{z}}^2+243.669 L_{\bar{z}}\right)+34.7654 L_{\bar{z}}^3+5.09037 L_{\bar{z}}^2-1826.42 L_{\bar{z}}+\bar{z} \left(-42.6667 L_{\bar{z}}^3+133.837 L_{\bar{z}}^2+2290.36 L_{\bar{z}}+473.618\right)-23.3915 z^6+335.792 z^5-11339.4 z^4-583.13
\,,
\end{dmath}

\begin{dmath}[style={\small},compact]
\delta C^{(2)}_{\bar q q}(z) =
z^3 \left(48.286 L_z^3-13.4154 L_z^2+671.421 L_z-392.838\right)+z^2 \left(-1.24024 L_z^3+25.0531 L_z^2+243.89 L_z+653.385\right)+z \left(2.66667 L_z^3-0.888889 L_z+6.41883\right)+0.888889 L_z-1.49899 z^6+20.8743 z^5-286.341 z^4-0.444444 \bar{z}
\,,
\end{dmath}

\begin{dmath}[style={\small},compact]
\delta C^{(3)}_{\bar q q}(z) =
N_f \left(z^3 \left(37.0364 L_z^4-346.932 L_z^3+1818.28 L_z^2-7637.09 L_z+11385.8\right)+z^2 \left(-3.3539 L_z^4-52.2866 L_z^3-733.542 L_z^2-4732.76 L_z-11873.\right)+z \left(1.54733 L_z^4-5.48697 L_z^3-3.1357 L_z^2-2.86412 L_z-25.5883\right)+0.790123 L_z^2-1.77778 L_z+\bar{z}^3 \left(-2.41513 L_{\bar{z}}^2-2.27567 L_{\bar{z}}\right)+\bar{z}^2 \left(0.096473 L_{\bar{z}}^2+3.48785 L_{\bar{z}}\right)+\bar{z} \left(0.197531 L_{\bar{z}}^2+1.0535 L_{\bar{z}}+0.92562\right)+8.42153 z^6-80.4256 z^5+585.108 z^4-0.254789\right)+z^3 \left(-1978.47 L_z^5+2860.25 L_z^4-82483.7 L_z^3+98475.2 L_z^2-514733. L_z-98159.7\right)+z^2 \left(37.7554 L_z^5+1076.04 L_z^4+14126.6 L_z^3+94540.7 L_z^2+239007. L_z-89211.8\right)+z \left(-8.17778 L_z^5-7.95062 L_z^4+244.41 L_z^3+90.8704 L_z^2-547.94 L_z-606.336\right)-7.11111 L_z^2+42.9687 L_z+\bar{z}^3 \left(53.1005 L_{\bar{z}}^2+66.5415 L_{\bar{z}}\right)+\bar{z}^2 \left(-1.30965 L_{\bar{z}}^2-58.0375 L_{\bar{z}}\right)+\bar{z} \left(-3.13186 L_{\bar{z}}^2-19.7773 L_{\bar{z}}-16.7318\right)-59.6703 z^6-2084.85 z^5+190130. z^4-8.07101
\,.
\end{dmath}

\section{Conclusion}
\label{sec:conclusion}
In this work we presented the first N$^3$LO twist-2 matching coefficients for TMD quark transversity and extracted the complete NNLO DGLAP splitting functions in both space-like and time-like regimes. We verified that our results for NNLO space-like  splitting functions are consistent with those in the literatures~\cite{Vogelsang:1997ak,Mikhailov:2008my,Blumlein:2021enk}.  We rederived the NNLO TMD transversity PDFs and FFs, noting minor discrepancies compared to the work of Ref.~\cite{Gutierrez-Reyes:2018iod}, whose origin will be investigated in future work.
Our results place transversity on the same theoretical footing as unpolarized and helicity sectors and provide key inputs for precision study of SIDIS azimuthal asymmetries at the forthcoming EIC.

\acknowledgments
I thank Hua Xing Zhu for encouraging me to look into this problem and Tong-Zhi Yang for collaboration in the early stages of the project.

\appendix

\section{QCD Beta Function}
\label{sec:beta}

The QCD beta function is defined as
\begin{equation}
\frac{d\alpha_s}{d\ln\mu} = \beta(\alpha_s) = -2\alpha_s \sum_{n=0}^\infty \left( \frac{\alpha_s}{4 \pi} \right)^{n+1} \, \beta_n \, ,
\end{equation}
with~\cite{Baikov:2016tgj}
\begin{align}
\beta_0 &= \frac{11}{3} C_A - \frac{4}{3} T_F N_f \, , \nn
\\
\beta_1 &= \frac{34}{3} C_A^2 - \frac{20}{3} C_A T_F N_f - 4 C_F T_F N_f \, ,\nn
\\
\beta_2 &= \left(\frac{158 C_A}{27}+\frac{44 C_F}{9}\right) N_f^2 T_F^2 +\left(-\frac{205 C_A
   C_F}{9}-\frac{1415 C_A^2}{27}+2 C_F^2\right) N_f T_F  +\frac{2857 C_A^3}{54}\,.
\end{align}

\section{Anomalous dimensions}
\label{sec:AD}

For all the anomalous dimensions entering the renormalization group equations of various TMD functions, we define the perturbative expansion in $\alpha_s$ according to
\begin{equation}
\gamma(\alpha_s) = \sum_{n=0}^\infty \left( \frac{\alpha_s}{4 \pi} \right)^{n+1} \, \gamma_n \, ,
\end{equation}
where the coefficients for quark  are given by
\begin{align}
\Gcusp_{0} =& 4 C_F\,, \nn
\\
\Gcusp_{1} =&  \left(\frac{268}{9}-8 
                 \zeta_2\right) C_A C_F -\frac{80 C_F T_F N_f}{9}\,, \nn
\\
\Gcusp_{2} =&\bigg[ \left(\frac{320 \zeta _2}{9}-\frac{224 \zeta _3}{3}-\frac{1672}{27}\right) C_A
   C_F+\left(64 \zeta _3-\frac{220}{3}\right) C_F^2\bigg] N_f T_F  \nn
   \\
 +&\left(-\frac{1072 \zeta
   _2}{9}+\frac{88 \zeta _3}{3}+88 \zeta _4+\frac{490}{3}\right) C_A^2 C_F  -\frac{64}{27} C_F
   N_f^2 T_F^2\,,\nn
     \\
\gamma^S_0 =& 0 \, , \nn
\\
\gamma^S_1 =& \left[ \left( -\frac{404}{27} + \frac{11\zeta_2}{3} + 14\zeta_3 \right) C_A   + \left( \frac{112}{27} - \frac{4 \zeta_2}{3} \right)T_F N_f   \right]  C_F   \,, \nn
\\
\gamma^S_2  =&\left(-\frac{88}{3} \zeta
   _3 \zeta _2+\frac{6325 \zeta _2}{81}+\frac{658 \zeta _3}{3}-88 \zeta
   _4-96 \zeta _5-\frac{136781}{1458}\right) C_A^2 C_F
+\left(\frac{80\zeta _2}{27}-\frac{224 \zeta _3}{27}\right.
\nn\\
+&\left.\frac{4160}{729}\right) C_FN_f^2 T_F^2\nn
    + \left(-\frac{2828 \zeta _2}{81}-\frac{728 \zeta _3}{27}+48 \zeta
   _4+\frac{11842}{729}\right) C_A C_F N_f T_F
   \nn\\
   +&\left(-4 \zeta _2-\frac{304 \zeta _3}{9}-16 \zeta
   _4+\frac{1711}{27}\right) C_F^2 N_f T_F\,.\nn
\nn\\
   \gamma^R_0 = &0 \, , \nn
\\
\gamma^R_1 = & \left[ \left( -\frac{404}{27} + 14\zeta_3 \right) C_A  +  \frac{112}{27} T_F N_f \right] C_F  \,, \nn 
\\
\gamma^R_2 =&\bigg[\left(-\frac{824 \zeta _2}{81}-\frac{904 \zeta _3}{27}+\frac{20 \zeta
   _4}{3}+\frac{62626}{729}\right) C_A N_f T_F 
   +\left(-\frac{88}{3} \zeta _3 \zeta
   _2+\frac{3196 \zeta _2}{81}+\frac{6164 \zeta _3}{27}\right.
   \nn\\
   +&\left.\frac{77 \zeta _4}{3}-96 \zeta_5
   -\frac{297029}{1458}\right)C_A^2 
   + \left(-\frac{304 \zeta _3}{9}-16 \zeta_4+\frac{1711}{27}\right)  C_F N_f T_F 
   +\left(-\frac{64 \zeta_3}{9}\right.
   \nn\\
   -&\left.\frac{3712}{729}\right) N_f^2 T_F^2 \bigg] C_F \,.
\end{align}
Since cusp and  soft and rapidity anomalous dimensions exhibit Casimir scaling,
 the corresponding anomalous dimensions for gluon could be obtained by multiplying in above with $C_A/C_F$.

The beam anomalous dimensions do not exhibits Casimir scaling, thus should be list separately.
The beam anomalous dimensions  for quark are
 \begin{align}
\gamma^B_0 =& 3C_F \, , \nn
\\
\gamma^B_1 =&  \left[  \left( \frac{3}{2} - 12\zeta_2 + 24\zeta_3 \right) C_F + \left( \frac{17}{6} + \frac{44 \zeta_2}{3} - 12\zeta_3 \right)  C_A 
 + \left( -\frac{2}{3} - \frac{16\zeta_2}{3} \right) T_F N_f  \right] C_F  , \nn
\\
\gamma^B_2 = &
 \bigg[ \left(-\frac{2672 \zeta _2}{27}+\frac{400 \zeta _3}{9}+4
   \zeta _4+40\right) C_A C_F+\left(\frac{40 \zeta _2}{3}-\frac{272 \zeta
   _3}{3}+\frac{232 \zeta _4}{3}-46\right) C_F^2\bigg] N_f T_F \nn 
\\   
  +& \left(16 \zeta _3
   \zeta _2-\frac{410 \zeta _2}{3}+\frac{844 \zeta _3}{3}-\frac{494 \zeta
   _4}{3}+120 \zeta _5+\frac{151}{4}\right) C_A C_F^2
   +\left(\frac{320 \zeta _2}{27}-\frac{64 \zeta _3}{9}-\frac{68}{9}\right)
   \nn\\
  \times&  C_F N_f^2T_F^2
  + \left(\frac{4496
   \zeta _2}{27}-\frac{1552 \zeta _3}{9}-5 \zeta _4+40 \zeta _5-\frac{1657}{36}\right) C_A^2 C_F
   +\bigg(-32 \zeta _3 \zeta _2+18 \zeta _2
   \nn\\
   +&68 \zeta _3+144 \zeta_4-240 \zeta _5+\frac{29}{2}\bigg) C_F^3\,.
\end{align}
The beam anomalous dimensions  for gluon are
 \begin{align}
\gamma^B_0 =& \frac{11}{3} C_A - \frac{4}{3} T_F N_f \,, \nn
\\
\gamma^B_1 =& C_A^2 \left( \frac{32}{3}+ 12 \zeta_3 \right) + \left(  -\frac{16}{3} C_A -  4 C_F \right) N_f T_F   \,, \nn
\\
\gamma^B_2 =& C_A^3\left(-80\zeta_5-16\zeta_3\zeta_2+\frac{55}{3}\zeta_4+\frac{536}{3}\zeta_3+\frac{8}{3}\zeta_2+\frac{79}{2}\right)
\nn\\
+&C_A^2 N_f T_F\left(-\frac{20}{3}\zeta_4-\frac{160}{3}\zeta_3-\frac{16}{3}\zeta_2-\frac{233}{9}\right)
+\frac{58}{9}C_A N_f^2 T_F^2
-\frac{241}{9}C_A C_F N_f T_F
\nn\\
+&2 C_F^2 N_f T_F +\frac{44}{9}C_F N_f^2 T_F^2\,.
\end{align}
The cusp anomalous dimension $\Gamma^{\text{cusp}}$ can be found in \cite{Moch:2004pa}. 
 The beam anomalous dimension $\gamma^B$ is related to the soft anomalous dimension $\gamma^S$~\cite{Li:2014afw}
  and  the hard anomalous dimensions $\gamma^H$~\cite{Moch:2005tm,Gehrmann:2010ue,Becher:2009qa} by renormalization group invariance condition $\gamma^B = \gamma^S - \gamma^H$.
 The rapidity anomalous dimension $\gamma^R$ can be found in \cite{Li:2016ctv,Vladimirov:2016dll}. 
 Note that the normalization here differ from those in \cite{Li:2016ctv} by a factor of $1/2$. 

\section{Renormalization Constants}
\label{sec:RC}
The following constants are needed for the renormalization of zero-bin subtracted~\cite{Manohar:2006nz} TMD PDFs through N$^3$LO, see e.g. Ref.~\cite{Luo:2019hmp,Luo:2019bmw}. 
The first three-order corrections to $Z^B $ and $Z^S$ are 
\begin{align}
\label{eqZqZs}
Z^B_1 =& \frac{1}{2\epsilon} \left(2 \gamma^B_0 -\Gamma_0^{\text{cusp}} L_Q \right) \,, \nonumber \\
Z^B_2 =& \frac{1}{8 \epsilon^2} \bigg( ( \Gamma_0^{\text{cusp}} L_Q - 2 \gamma^B_0)^2 + 2 \beta_0 (  \Gamma_0^{\text{cusp}} L_Q - 2 \gamma^B_0)    \bigg)
 + \frac{1}{4\epsilon} \left( 2 \gamma^B_1 - \Gamma_1^{\text{cusp}} L_Q \right) \,, \nonumber \\
Z^B_3  =& \frac{1}{48 \epsilon^3}  \left( 2 \gamma^B_0 -\Gamma^{\text{cusp}}_0 L_Q \right) \biggl( 8 \beta_0^2 + 6 \beta_0 \left( -2 \gamma^B_0 + \Gamma^{\text{cusp}}_0 L_Q \right) 
+ \left( -2 \gamma^B_0 + \Gamma^{\text{cusp}}_0 L_Q \right)^2 \biggl) \nn \\ 
+& \frac{1}{24 \epsilon^2} \biggl( \beta_1 \left(-8 \gamma^B_0 + 4  \Gamma^{\text{cusp}}_0 L_Q \right) + \left(4 \beta_0 - 6 \gamma^B_0 + 3 \Gamma^{\text{cusp}}_0 L_Q \right) \left( -2 \gamma^B_1 + \Gamma^{\text{cusp}}_1 L_Q \right)  \biggl) 
 \nn  \\    
  +& \frac{1}{6 \epsilon} \biggl(  2 \gamma^B_2 -  \Gamma^{\text{cusp}}_2 L_Q  \biggl) 
  \nn\\  
Z^S_1 =& \frac{1}{\epsilon^2} \Gamma^{\text{cusp}}_0  +  \frac{1}{\epsilon} \left( -2 \gamma^S_0 - \Gamma_0^{\text{cusp}} L_\nu \right) \,,\nonumber \\
Z^S_2 =& \frac{1}{2 \epsilon^4} (\Gamma^{\text{cusp}}_0)^2 - \frac{1}{4 \epsilon^3} \bigg(\Gamma^{\text{cusp}}_0 (3 \beta_0 + 8 \gamma^S_0)+4( \Gamma^{\text{cusp}}_0)^2 L_\nu\bigg)   - \frac{1}{2 \epsilon} \left( 2 \gamma^S_1 +  \Gamma^{\text{cusp}}_1 L_\nu \right) \nonumber \\
+& \frac{1}{4 \epsilon^2} \bigg(\Gamma^{\text{cusp}}_1 + 2 ( 2 \gamma^S_0 + \Gamma^{\text{cusp}}_0 L_\nu ) ( \beta_0 + 2 \gamma^S_0 + \Gamma^{\text{cusp}}_0 L_\nu) \bigg) \,, \nn \\
Z^S_3 =& \frac{ 1}{6 \epsilon^6} \left(\Gamma^{\text{cusp}}_0\right)^3  - \frac{1}{4 \epsilon^5} \left(\Gamma^{\text{cusp}}_0 \right)^2 \left(  3 \beta_0 + 4 \gamma^S_0 + 2 \Gamma^{\text{cusp}}_0 L_\nu \right) + \frac{1}{36 \epsilon^4 } \Gamma^{\text{cusp}}_0 \bigg( 22 \beta_0^2 + 45 \beta_0 \left(2 \gamma^S_0 + \Gamma^{\text{cusp}}_0 L_\nu \right)  \nn \\
+& 9 \left( \Gamma^{\text{cusp}}_1 + 2 \left( 2 \gamma^S_0+ \Gamma^{\text{cusp}}_0 L_\nu\right)^2  \right)   \biggl)  + \frac{1}{36 \epsilon^3} \biggl( -16 \beta_1 \Gamma^{\text{cusp}}_0 - 12 \beta_0^2 \left( 2 \gamma^S_0 + \Gamma^{\text{cusp}}_0 L_\nu \right) \nn \\
 -& 2 \beta_0 \left( 5 \Gamma^{\text{cusp}}_1 + 9 \left( 2 \gamma^S_0 + \Gamma^{\text{cusp}}_0 L_\nu \right)^2 \right) - 3 \bigg[ \Gamma^{\text{cusp}}_1 \left(6 \gamma^S_0 + 9 \Gamma^{\text{cusp}}_0 L_\nu \right)  \nn \\
+& 2 \left( 8 \left( \gamma^S_0\right)^3 + 6 \Gamma^{\text{cusp}}_0 \gamma^S_1 + 12 \Gamma^{\text{cusp}}_0  \left(\gamma^S_0\right)^2 L_\nu + 6 \left(\Gamma^{\text{cusp}}_0\right)^2 \gamma^S_0 L_\nu^2 + \left( \Gamma^{\text{cusp}}_0 \right)^3 L_\nu^3 \right) \bigg]    \biggl) \nn \\
 +& \frac{1}{18 \epsilon^2} \biggl(  2 \Gamma^{\text{cusp}}_2 + 3 \left( 2 \beta_1 \left( 2 \gamma^S_0 + \Gamma^{\text{cusp}}_0 L_\nu \right) + \left( 2 \beta_0 + 6 \gamma^S_0 + 3 \Gamma^{\text{cusp}}_0 L_\nu \right) \left( 2 \gamma^S_1 + \Gamma^{\text{cusp}}_1 L_\nu \right) \right)    \biggl) 
 \nn\\
 - & \frac{2 \gamma^S_2 + \Gamma^{\text{cusp}}_2 L_\nu}{3 \epsilon} \,.
\end{align}
Keep in mind that the  anomalous dimensions appeared above depends on the flavor, they should be replaced by the corresponding values in Sec~\ref{sec:AD}.
We also remind the reader that the renormalization constants are formally identical for TMD PDFs and TMD FFs,
the logarithms appeared above should be replaced by their corresponding values in each case,
and we have
\begin{align}
\label{eq:LdefinitionALL}
 L_\perp = \ln \frac{b_T^2 \mu^2}{b_0^2} ,  \quad L_\nu = \ln \frac{\nu^2}{\mu^2} \,,
\end{align}
with $b_0 =2 \, e^{- \gamma_E}$ for both TMD PDFs and TMD FFs.
On the other hand, we have for TMD PDFs
\begin{align}
\label{eq:LdefinitionSLQ}
 L_Q  = 2 \ln \frac{x \, P_+}{\nu} \,,
\end{align}
while for TMD FFs,
\begin{align}
\label{eq:LdefinitionTLQ}
L_Q = 2 \ln \frac{P_+}{ z \, \nu}.
\end{align}

\bibliographystyle{JHEP}
\bibliography{transversity_TMD}

\providecommand{\href}[2]{#2}\begingroup\raggedright\begin{thebibliography}{100}

\bibitem{Collins:2011zzd}
J.~Collins, {\it {Foundations of perturbative QCD}},  {\em Camb. Monogr. Part.
  Phys. Nucl. Phys. Cosmol.} {\bf 32} (2011) 1--624.

\bibitem{Ralston:1979ys}
J.~P. Ralston and D.~E. Soper, {\it {Production of Dimuons from High-Energy
  Polarized Proton Proton Collisions}},  {\em Nucl. Phys. B} {\bf 152} (1979)
  109.

\bibitem{Jaffe:1991kp}
R.~L. Jaffe and X.-D. Ji, {\it {Chiral odd parton distributions and polarized
  Drell-Yan}},  {\em Phys. Rev. Lett.} {\bf 67} (1991) 552--555.

\bibitem{Jaffe:1991ra}
R.~L. Jaffe and X.-D. Ji, {\it {Chiral odd parton distributions and Drell-Yan
  processes}},  {\em Nucl. Phys. B} {\bf 375} (1992) 527--560.

\bibitem{Ji:1992ev}
X.-D. Ji, {\it {Probing the nucleon's transversity distribution in polarized p
  p, p anti-p, and pi p collisions}},  {\em Phys. Lett. B} {\bf 284} (1992)
  137--143.

\bibitem{Jaffe:1993xb}
R.~L. Jaffe and X.-D. Ji, {\it {Novel quark fragmentation functions and the
  nucleon's transversity distribution}},  {\em Phys. Rev. Lett.} {\bf 71}
  (1993) 2547--2550, [\href{http://arxiv.org/abs/hep-ph/9307329}{{\tt
  hep-ph/9307329}}].

\bibitem{Courtoy:2015haa}
A.~Courtoy, S.~Bae{\ss}ler, M.~Gonz{\'a}lez-Alonso, and S.~Liuti, {\it
  {Beyond-Standard-Model Tensor Interaction and Hadron Phenomenology}},  {\em
  Phys. Rev. Lett.} {\bf 115} (2015) 162001,
  [\href{http://arxiv.org/abs/1503.06814}{{\tt arXiv:1503.06814}}].

\bibitem{Cirigliano:2013xha}
V.~Cirigliano, S.~Gardner, and B.~Holstein, {\it {Beta Decays and Non-Standard
  Interactions in the LHC Era}},  {\em Prog. Part. Nucl. Phys.} {\bf 71} (2013)
  93--118, [\href{http://arxiv.org/abs/1303.6953}{{\tt arXiv:1303.6953}}].

\bibitem{Yamanaka:2017mef}
N.~Yamanaka, B.~K. Sahoo, N.~Yoshinaga, T.~Sato, K.~Asahi, and B.~P. Das, {\it
  {Probing exotic phenomena at the interface of nuclear and particle physics
  with the electric dipole moments of diamagnetic atoms: A unique window to
  hadronic and semi-leptonic CP violation}},  {\em Eur. Phys. J. A} {\bf 53}
  (2017), no.~3 54, [\href{http://arxiv.org/abs/1703.01570}{{\tt
  arXiv:1703.01570}}].

\bibitem{Efremov:1992pe}
A.~V. Efremov, L.~Mankiewicz, and N.~A. Tornqvist, {\it {Jet handedness as a
  measure of quark and gluon polarization}},  {\em Phys. Lett. B} {\bf 284}
  (1992) 394--400.

\bibitem{Collins:1993kq}
J.~C. Collins, S.~F. Heppelmann, and G.~A. Ladinsky, {\it {Measuring
  transversity densities in singly polarized hadron hadron and lepton - hadron
  collisions}},  {\em Nucl. Phys. B} {\bf 420} (1994) 565--582,
  [\href{http://arxiv.org/abs/hep-ph/9305309}{{\tt hep-ph/9305309}}].

\bibitem{Jaffe:1997hf}
R.~L. Jaffe, X.-m. Jin, and J.~Tang, {\it {Interference fragmentation functions
  and the nucleon's transversity}},  {\em Phys. Rev. Lett.} {\bf 80} (1998)
  1166--1169, [\href{http://arxiv.org/abs/hep-ph/9709322}{{\tt
  hep-ph/9709322}}].

\bibitem{Qiu:1998ia}
J.-w. Qiu and G.~F. Sterman, {\it {Single transverse spin asymmetries in
  hadronic pion production}},  {\em Phys. Rev. D} {\bf 59} (1999) 014004,
  [\href{http://arxiv.org/abs/hep-ph/9806356}{{\tt hep-ph/9806356}}].

\bibitem{Bianconi:1999cd}
A.~Bianconi, S.~Boffi, R.~Jakob, and M.~Radici, {\it {Two hadron interference
  fragmentation functions. Part 1. General framework}},  {\em Phys. Rev. D}
  {\bf 62} (2000) 034008, [\href{http://arxiv.org/abs/hep-ph/9907475}{{\tt
  hep-ph/9907475}}].

\bibitem{Radici:2001na}
M.~Radici, R.~Jakob, and A.~Bianconi, {\it {Accessing transversity with
  interference fragmentation functions}},  {\em Phys. Rev. D} {\bf 65} (2002)
  074031, [\href{http://arxiv.org/abs/hep-ph/0110252}{{\tt hep-ph/0110252}}].

\bibitem{Barone:2001sp}
V.~Barone, A.~Drago, and P.~G. Ratcliffe, {\it {Transverse polarisation of
  quarks in hadrons}},  {\em Phys. Rept.} {\bf 359} (2002) 1--168,
  [\href{http://arxiv.org/abs/hep-ph/0104283}{{\tt hep-ph/0104283}}].

\bibitem{Bacchetta:2002ux}
A.~Bacchetta and M.~Radici, {\it {Partial wave analysis of two hadron
  fragmentation functions}},  {\em Phys. Rev. D} {\bf 67} (2003) 094002,
  [\href{http://arxiv.org/abs/hep-ph/0212300}{{\tt hep-ph/0212300}}].

\bibitem{Collins:1992kk}
J.~C. Collins, {\it {Fragmentation of transversely polarized quarks probed in
  transverse momentum distributions}},  {\em Nucl. Phys. B} {\bf 396} (1993)
  161--182, [\href{http://arxiv.org/abs/hep-ph/9208213}{{\tt hep-ph/9208213}}].

\bibitem{Anselmino:2015sxa}
M.~Anselmino, M.~Boglione, U.~D'Alesio, J.~O. Gonzalez~Hernandez, S.~Melis,
  F.~Murgia, and A.~Prokudin, {\it {Collins functions for pions from SIDIS and
  new $e^+e^-$ data: a first glance at their transverse momentum dependence}},
  {\em Phys. Rev. D} {\bf 92} (2015), no.~11 114023,
  [\href{http://arxiv.org/abs/1510.05389}{{\tt arXiv:1510.05389}}].

\bibitem{Kang:2015msa}
Z.-B. Kang, A.~Prokudin, P.~Sun, and F.~Yuan, {\it {Extraction of Quark
  Transversity Distribution and Collins Fragmentation Functions with QCD
  Evolution}},  {\em Phys. Rev. D} {\bf 93} (2016), no.~1 014009,
  [\href{http://arxiv.org/abs/1505.05589}{{\tt arXiv:1505.05589}}].

\bibitem{Lin:2017stx}
H.-W. Lin, W.~Melnitchouk, A.~Prokudin, N.~Sato, and H.~Shows, {\it {First
  Monte Carlo Global Analysis of Nucleon Transversity with Lattice QCD
  Constraints}},  {\em Phys. Rev. Lett.} {\bf 120} (2018), no.~15 152502,
  [\href{http://arxiv.org/abs/1710.09858}{{\tt arXiv:1710.09858}}].

\bibitem{DAlesio:2020vtw}
U.~D'Alesio, C.~Flore, and A.~Prokudin, {\it {Role of the Soffer bound in
  determination of transversity and the tensor charge}},  {\em Phys. Lett. B}
  {\bf 803} (2020) 135347, [\href{http://arxiv.org/abs/2001.01573}{{\tt
  arXiv:2001.01573}}].

\bibitem{Cammarota:2020qcw}
{\bf Jefferson Lab Angular Momentum} Collaboration, J.~Cammarota, L.~Gamberg,
  Z.-B. Kang, J.~A. Miller, D.~Pitonyak, A.~Prokudin, T.~C. Rogers, and
  N.~Sato, {\it {Origin of single transverse-spin asymmetries in high-energy
  collisions}},  {\em Phys. Rev. D} {\bf 102} (2020), no.~5 054002,
  [\href{http://arxiv.org/abs/2002.08384}{{\tt arXiv:2002.08384}}].

\bibitem{Gamberg:2022kdb}
{\bf Jefferson Lab Angular Momentum (JAM), Jefferson Lab Angular Momentum}
  Collaboration, L.~Gamberg, M.~Malda, J.~A. Miller, D.~Pitonyak, A.~Prokudin,
  and N.~Sato, {\it {Updated QCD global analysis of single transverse-spin
  asymmetries: Extracting H{\textasciitilde}, and the role of the Soffer bound
  and lattice QCD}},  {\em Phys. Rev. D} {\bf 106} (2022), no.~3 034014,
  [\href{http://arxiv.org/abs/2205.00999}{{\tt arXiv:2205.00999}}].

\bibitem{Boglione:2024dal}
M.~Boglione, U.~D'Alesio, C.~Flore, J.~O. Gonzalez-Hernandez, F.~Murgia, and
  A.~Prokudin, {\it {Simultaneous reweighting of Transverse Momentum Dependent
  distributions}},  {\em Phys. Lett. B} {\bf 854} (2024) 138712,
  [\href{http://arxiv.org/abs/2402.12322}{{\tt arXiv:2402.12322}}].

\bibitem{Radici:2018iag}
M.~Radici and A.~Bacchetta, {\it {First Extraction of Transversity from a
  Global Analysis of Electron-Proton and Proton-Proton Data}},  {\em Phys. Rev.
  Lett.} {\bf 120} (2018), no.~19 192001,
  [\href{http://arxiv.org/abs/1802.05212}{{\tt arXiv:1802.05212}}].

\bibitem{Benel:2019mcq}
J.~Benel, A.~Courtoy, and R.~Ferro-Hernandez, {\it {A constrained fit of the
  valence transversity distributions from dihadron production}},  {\em Eur.
  Phys. J. C} {\bf 80} (2020), no.~5 465,
  [\href{http://arxiv.org/abs/1912.03289}{{\tt arXiv:1912.03289}}].

\bibitem{Cocuzza:2023oam}
{\bf JAM} Collaboration, C.~Cocuzza, A.~Metz, D.~Pitonyak, A.~Prokudin,
  N.~Sato, and R.~Seidl, {\it {Transversity Distributions and Tensor Charges of
  the Nucleon: Extraction from Dihadron Production and Their Universal
  Nature}},  {\em Phys. Rev. Lett.} {\bf 132} (2024), no.~9 091901,
  [\href{http://arxiv.org/abs/2306.12998}{{\tt arXiv:2306.12998}}].

\bibitem{HERMES:2004mhh}
{\bf HERMES} Collaboration, A.~Airapetian et~al., {\it {Single-spin asymmetries
  in semi-inclusive deep-inelastic scattering on a transversely polarized
  hydrogen target}},  {\em Phys. Rev. Lett.} {\bf 94} (2005) 012002,
  [\href{http://arxiv.org/abs/hep-ex/0408013}{{\tt hep-ex/0408013}}].

\bibitem{Yuan:2007nd}
F.~Yuan, {\it {Azimuthal asymmetric distribution of hadrons inside a jet at
  hadron collider}},  {\em Phys. Rev. Lett.} {\bf 100} (2008) 032003,
  [\href{http://arxiv.org/abs/0709.3272}{{\tt arXiv:0709.3272}}].

\bibitem{Burkardt:2008jw}
M.~Burkardt, C.~A. Miller, and W.~D. Nowak, {\it {Spin-polarized high-energy
  scattering of charged leptons on nucleons}},  {\em Rept. Prog. Phys.} {\bf
  73} (2010) 016201, [\href{http://arxiv.org/abs/0812.2208}{{\tt
  arXiv:0812.2208}}].

\bibitem{Anselmino:2007fs}
M.~Anselmino, M.~Boglione, U.~D'Alesio, A.~Kotzinian, F.~Murgia, A.~Prokudin,
  and C.~Turk, {\it {Transversity and Collins functions from SIDIS and e+ e-
  data}},  {\em Phys. Rev. D} {\bf 75} (2007) 054032,
  [\href{http://arxiv.org/abs/hep-ph/0701006}{{\tt hep-ph/0701006}}].

\bibitem{COMPASS:2008isr}
{\bf COMPASS} Collaboration, M.~Alekseev et~al., {\it {Collins and Sivers
  asymmetries for pions and kaons in muon-deuteron DIS}},  {\em Phys. Lett. B}
  {\bf 673} (2009) 127--135, [\href{http://arxiv.org/abs/0802.2160}{{\tt
  arXiv:0802.2160}}].

\bibitem{HERMES:2010mmo}
{\bf HERMES} Collaboration, A.~Airapetian et~al., {\it {Effects of transversity
  in deep-inelastic scattering by polarized protons}},  {\em Phys. Lett. B}
  {\bf 693} (2010) 11--16, [\href{http://arxiv.org/abs/1006.4221}{{\tt
  arXiv:1006.4221}}].

\bibitem{COMPASS:2010hbb}
{\bf COMPASS} Collaboration, M.~G. Alekseev et~al., {\it {Measurement of the
  Collins and Sivers asymmetries on transversely polarised protons}},  {\em
  Phys. Lett. B} {\bf 692} (2010) 240--246,
  [\href{http://arxiv.org/abs/1005.5609}{{\tt arXiv:1005.5609}}].

\bibitem{Courtoy:2012ry}
A.~Courtoy, A.~Bacchetta, M.~Radici, and A.~Bianconi, {\it {First extraction of
  Interference Fragmentation Functions from $e^+e^-$ data}},  {\em Phys. Rev.
  D} {\bf 85} (2012) 114023, [\href{http://arxiv.org/abs/1202.0323}{{\tt
  arXiv:1202.0323}}].

\bibitem{Martin:2014wua}
A.~Martin, F.~Bradamante, and V.~Barone, {\it {Extracting the transversity
  distributions from single-hadron and dihadron production}},  {\em Phys. Rev.
  D} {\bf 91} (2015), no.~1 014034, [\href{http://arxiv.org/abs/1412.5946}{{\tt
  arXiv:1412.5946}}].

\bibitem{COMPASS:2014bze}
{\bf COMPASS} Collaboration, C.~Adolph et~al., {\it {Collins and Sivers
  asymmetries in muonproduction of pions and kaons off transversely polarised
  protons}},  {\em Phys. Lett. B} {\bf 744} (2015) 250--259,
  [\href{http://arxiv.org/abs/1408.4405}{{\tt arXiv:1408.4405}}].

\bibitem{STAR:2017wsi}
{\bf STAR} Collaboration, L.~Adamczyk et~al., {\it {Transverse spin-dependent
  azimuthal correlations of charged pion pairs measured in p$^\uparrow$+p
  collisions at $\sqrt{s}$ = 500 GeV}},  {\em Phys. Lett. B} {\bf 780} (2018)
  332--339, [\href{http://arxiv.org/abs/1710.10215}{{\tt arXiv:1710.10215}}].

\bibitem{Anselmino:2013vqa}
M.~Anselmino, M.~Boglione, U.~D'Alesio, S.~Melis, F.~Murgia, and A.~Prokudin,
  {\it {Simultaneous extraction of transversity and Collins functions from new
  SIDIS and e+e- data}},  {\em Phys. Rev. D} {\bf 87} (2013) 094019,
  [\href{http://arxiv.org/abs/1303.3822}{{\tt arXiv:1303.3822}}].

\bibitem{Kang:2014zza}
Z.-B. Kang, A.~Prokudin, P.~Sun, and F.~Yuan, {\it {Nucleon tensor charge from
  Collins azimuthal asymmetry measurements}},  {\em Phys. Rev. D} {\bf 91}
  (2015) 071501, [\href{http://arxiv.org/abs/1410.4877}{{\tt
  arXiv:1410.4877}}].

\bibitem{Cocuzza:2023vqs}
{\bf Jefferson Lab Angular Momentum (JAM)} Collaboration, C.~Cocuzza, A.~Metz,
  D.~Pitonyak, A.~Prokudin, N.~Sato, and R.~Seidl, {\it {First simultaneous
  global QCD analysis of dihadron fragmentation functions and transversity
  parton distribution functions}},  {\em Phys. Rev. D} {\bf 109} (2024), no.~3
  034024, [\href{http://arxiv.org/abs/2308.14857}{{\tt arXiv:2308.14857}}].

\bibitem{COMPASS:2023vhr}
{\bf COMPASS} Collaboration, G.~D. Alexeev et~al., {\it {High-Statistics
  Measurement of Collins and Sivers Asymmetries for Transversely Polarized
  Deuterons}},  {\em Phys. Rev. Lett.} {\bf 133} (2024), no.~10 101903,
  [\href{http://arxiv.org/abs/2401.00309}{{\tt arXiv:2401.00309}}].

\bibitem{COMPASS:2023vqt}
{\bf COMPASS} Collaboration, G.~D. Alexeev et~al., {\it {Final COMPASS Results
  on the Transverse-Spin-Dependent Azimuthal Asymmetries in the Pion-Induced
  Drell-Yan Process}},  {\em Phys. Rev. Lett.} {\bf 133} (2024), no.~7 071902,
  [\href{http://arxiv.org/abs/2312.17379}{{\tt arXiv:2312.17379}}].

\bibitem{Zeng:2024gun}
C.~Zeng, H.~Dong, T.~Liue, P.~Sun, and Y.~Zhao, {\it {Global analysis of Sivers
  and Collins asymmetries within the TMD factorization}},
  \href{http://arxiv.org/abs/2412.18324}{{\tt arXiv:2412.18324}}.

\bibitem{STAR:2025xyp}
{\bf STAR} Collaboration, {\it {Energy Independence of the Collins Asymmetry in
  $p^{\uparrow}p$ Collisions}},  \href{http://arxiv.org/abs/2507.16355}{{\tt
  arXiv:2507.16355}}.

\bibitem{LHCspin:2025lvj}
{\bf LHCspin} Collaboration, A.~Accardi et~al., {\it {LHCspin: a Polarized Gas
  Target for LHC}},  \href{http://arxiv.org/abs/2504.16034}{{\tt
  arXiv:2504.16034}}.

\bibitem{Gao:2025evv}
M.-S. Gao, Z.-B. Kang, W.~Li, and D.~Y. Shao, {\it {Accessing nucleon
  transversity with one-point energy correlators}},
  \href{http://arxiv.org/abs/2509.15809}{{\tt arXiv:2509.15809}}.

\bibitem{DAlesio:2010sag}
U.~D'Alesio, F.~Murgia, and C.~Pisano, {\it {Azimuthal asymmetries for hadron
  distributions inside a jet in hadronic collisions}},  {\em Phys. Rev. D} {\bf
  83} (2011) 034021, [\href{http://arxiv.org/abs/1011.2692}{{\tt
  arXiv:1011.2692}}].

\bibitem{DAlesio:2025jmr}
U.~D'Alesio, C.~Flore, and M.~Zaccheddu, {\it {Collins function for pion-in-jet
  production in polarized pp collisions: a test of universality and
  factorization}},  {\em Phys. Lett. B} {\bf 871} (2025) 140024,
  [\href{http://arxiv.org/abs/2506.21959}{{\tt arXiv:2506.21959}}].

\bibitem{Cao:2025icu}
Q.-H. Cao, Z.~Yu, C.~P. Yuan, S.~Zhang, and H.~X. Zhu, {\it {Collins-type
  fragmentation energy correlator in semi-inclusive deep inelastic
  lepton-hadron scattering}},  \href{http://arxiv.org/abs/2509.18892}{{\tt
  arXiv:2509.18892}}.

\bibitem{Alexandrou:2018eet}
C.~Alexandrou, K.~Cichy, M.~Constantinou, K.~Jansen, A.~Scapellato, and
  F.~Steffens, {\it {Transversity parton distribution functions from lattice
  QCD}},  {\em Phys. Rev. D} {\bf 98} (2018), no.~9 091503,
  [\href{http://arxiv.org/abs/1807.00232}{{\tt arXiv:1807.00232}}].

\bibitem{Alexandrou:2016jqi}
C.~Alexandrou, K.~Cichy, M.~Constantinou, K.~Hadjiyiannakou, K.~Jansen,
  F.~Steffens, and C.~Wiese, {\it {Updated Lattice Results for Parton
  Distributions}},  {\em Phys. Rev. D} {\bf 96} (2017), no.~1 014513,
  [\href{http://arxiv.org/abs/1610.03689}{{\tt arXiv:1610.03689}}].

\bibitem{Chen:2016utp}
J.-W. Chen, S.~D. Cohen, X.~Ji, H.-W. Lin, and J.-H. Zhang, {\it {Nucleon
  Helicity and Transversity Parton Distributions from Lattice QCD}},  {\em
  Nucl. Phys. B} {\bf 911} (2016) 246--273,
  [\href{http://arxiv.org/abs/1603.06664}{{\tt arXiv:1603.06664}}].

\bibitem{HadStruc:2021qdf}
{\bf HadStruc} Collaboration, C.~Egerer et~al., {\it {Transversity parton
  distribution function of the nucleon using the pseudodistribution approach}},
   {\em Phys. Rev. D} {\bf 105} (2022), no.~3 034507,
  [\href{http://arxiv.org/abs/2111.01808}{{\tt arXiv:2111.01808}}].

\bibitem{LatticeParton:2022xsd}
{\bf Lattice Parton} Collaboration, F.~Yao et~al., {\it {Nucleon Transversity
  Distribution in the Continuum and Physical Mass Limit from Lattice QCD}},
  {\em Phys. Rev. Lett.} {\bf 131} (2023), no.~26 261901,
  [\href{http://arxiv.org/abs/2208.08008}{{\tt arXiv:2208.08008}}].

\bibitem{Gao:2023ktu}
X.~Gao, A.~D. Hanlon, S.~Mukherjee, P.~Petreczky, Q.~Shi, S.~Syritsyn, and
  Y.~Zhao, {\it {Transversity PDFs of the proton from lattice QCD with physical
  quark masses}},  {\em Phys. Rev. D} {\bf 109} (2024), no.~5 054506,
  [\href{http://arxiv.org/abs/2310.19047}{{\tt arXiv:2310.19047}}].

\bibitem{Aidala:2012mv}
C.~A. Aidala, S.~D. Bass, D.~Hasch, and G.~K. Mallot, {\it {The Spin Structure
  of the Nucleon}},  {\em Rev. Mod. Phys.} {\bf 85} (2013) 655--691,
  [\href{http://arxiv.org/abs/1209.2803}{{\tt arXiv:1209.2803}}].

\bibitem{Boussarie:2023izj}
R.~Boussarie et~al., {\it {TMD Handbook}},
  \href{http://arxiv.org/abs/2304.03302}{{\tt arXiv:2304.03302}}.

\bibitem{Luo:2019bmw}
M.-X. Luo, T.-Z. Yang, H.~X. Zhu, and Y.~J. Zhu, {\it {Transverse Parton
  Distribution and Fragmentation Functions at NNLO: the Gluon Case}},
  \href{http://arxiv.org/abs/1909.13820}{{\tt arXiv:1909.13820}}.

\bibitem{Luo:2020epw}
M.-x. Luo, T.-Z. Yang, H.~X. Zhu, and Y.~J. Zhu, {\it {Unpolarized Quark and
  Gluon TMD PDFs and FFs at N$^3$LO}},
  \href{http://arxiv.org/abs/2012.03256}{{\tt arXiv:2012.03256}}.

\bibitem{Ebert:2020qef}
M.~A. Ebert, B.~Mistlberger, and G.~Vita, {\it {TMD fragmentation functions at
  N$^{3}$LO}},  {\em JHEP} {\bf 07} (2021) 121,
  [\href{http://arxiv.org/abs/2012.07853}{{\tt arXiv:2012.07853}}].

\bibitem{Ebert:2020yqt}
M.~A. Ebert, B.~Mistlberger, and G.~Vita, {\it {Transverse momentum dependent
  PDFs at N$^3$LO}},  {\em JHEP} {\bf 09} (2020) 146,
  [\href{http://arxiv.org/abs/2006.05329}{{\tt arXiv:2006.05329}}].

\bibitem{Zhu:2025gts}
Y.~J. Zhu, {\it {The N$^3$LO Twist-2 Matching of Helicity TMDs and SIDIS
  $q_\ast$ Spectrum}},  \href{http://arxiv.org/abs/2509.01655}{{\tt
  arXiv:2509.01655}}.

\bibitem{Gutierrez-Reyes:2017glx}
D.~Guti{\'e}rrez-Reyes, I.~Scimemi, and A.~A. Vladimirov, {\it {Twist-2
  matching of transverse momentum dependent distributions}},  {\em Phys. Lett.
  B} {\bf 769} (2017) 84--89, [\href{http://arxiv.org/abs/1702.06558}{{\tt
  arXiv:1702.06558}}].

\bibitem{Gutierrez-Reyes:2018iod}
D.~Gutierrez-Reyes, I.~Scimemi, and A.~Vladimirov, {\it {Transverse momentum
  dependent transversely polarized distributions at
  next-to-next-to-leading-order}},  {\em JHEP} {\bf 07} (2018) 172,
  [\href{http://arxiv.org/abs/1805.07243}{{\tt arXiv:1805.07243}}].

\bibitem{She:2009jq}
J.~She, J.~Zhu, and B.-Q. Ma, {\it {Pretzelosity h(1T)**perpendicular and quark
  orbital angular momentum}},  {\em Phys. Rev. D} {\bf 79} (2009) 054008,
  [\href{http://arxiv.org/abs/0902.3718}{{\tt arXiv:0902.3718}}].

\bibitem{Avakian:2009jt}
H.~Avakian, A.~V. Efremov, P.~Schweitzer, O.~V. Teryaev, F.~Yuan, and
  P.~Zavada, {\it {Insights on non-perturbative aspects of TMDs from models}},
  {\em Mod. Phys. Lett. A} {\bf 24} (2009) 2995--3004,
  [\href{http://arxiv.org/abs/0910.3181}{{\tt arXiv:0910.3181}}].

\bibitem{Avakian:2010br}
H.~Avakian, A.~V. Efremov, P.~Schweitzer, and F.~Yuan, {\it {The transverse
  momentum dependent distribution functions in the bag model}},  {\em Phys.
  Rev. D} {\bf 81} (2010) 074035, [\href{http://arxiv.org/abs/1001.5467}{{\tt
  arXiv:1001.5467}}].

\bibitem{Li:2016axz}
Y.~Li, D.~Neill, and H.~X. Zhu, {\it {An Exponential Regulator for Rapidity
  Divergences}},  {\em Submitted to: Phys. Rev. D} (2016)
  [\href{http://arxiv.org/abs/1604.00392}{{\tt arXiv:1604.00392}}].

\bibitem{Blumlein:2021enk}
J.~Bl{\"u}mlein, P.~Marquard, C.~Schneider, and K.~Sch{\"o}nwald, {\it {The
  three-loop unpolarized and polarized non-singlet anomalous dimensions from
  off shell operator matrix elements}},  {\em Nucl. Phys. B} {\bf 971} (2021)
  115542, [\href{http://arxiv.org/abs/2107.06267}{{\tt arXiv:2107.06267}}].

\bibitem{Bauer:2000ew}
C.~W. Bauer, S.~Fleming, and M.~E. Luke, {\it {Summing Sudakov logarithms in B
  ---> X(s gamma) in effective field theory}},  {\em Phys. Rev.} {\bf D63}
  (2000) 014006, [\href{http://arxiv.org/abs/hep-ph/0005275}{{\tt
  hep-ph/0005275}}].

\bibitem{Bauer:2000yr}
C.~W. Bauer, S.~Fleming, D.~Pirjol, and I.~W. Stewart, {\it {An Effective field
  theory for collinear and soft gluons: Heavy to light decays}},  {\em Phys.
  Rev.} {\bf D63} (2001) 114020,
  [\href{http://arxiv.org/abs/hep-ph/0011336}{{\tt hep-ph/0011336}}].

\bibitem{Bauer:2001yt}
C.~W. Bauer, D.~Pirjol, and I.~W. Stewart, {\it {Soft collinear factorization
  in effective field theory}},  {\em Phys. Rev.} {\bf D65} (2002) 054022,
  [\href{http://arxiv.org/abs/hep-ph/0109045}{{\tt hep-ph/0109045}}].

\bibitem{Bauer:2002nz}
C.~W. Bauer, S.~Fleming, D.~Pirjol, I.~Z. Rothstein, and I.~W. Stewart, {\it
  {Hard scattering factorization from effective field theory}},  {\em Phys.
  Rev.} {\bf D66} (2002) 014017,
  [\href{http://arxiv.org/abs/hep-ph/0202088}{{\tt hep-ph/0202088}}].

\bibitem{Beneke:2002ph}
M.~Beneke, A.~P. Chapovsky, M.~Diehl, and T.~Feldmann, {\it {Soft collinear
  effective theory and heavy to light currents beyond leading power}},  {\em
  Nucl. Phys.} {\bf B643} (2002) 431--476,
  [\href{http://arxiv.org/abs/hep-ph/0206152}{{\tt hep-ph/0206152}}].

\bibitem{Bauer:2001ct}
C.~W. Bauer and I.~W. Stewart, {\it {Invariant operators in collinear effective
  theory}},  {\em Phys. Lett.} {\bf B516} (2001) 134--142,
  [\href{http://arxiv.org/abs/hep-ph/0107001}{{\tt hep-ph/0107001}}].

\bibitem{Goeke:2005hb}
K.~Goeke, A.~Metz, and M.~Schlegel, {\it {Parameterization of the quark-quark
  correlator of a spin-1/2 hadron}},  {\em Phys. Lett. B} {\bf 618} (2005)
  90--96, [\href{http://arxiv.org/abs/hep-ph/0504130}{{\tt hep-ph/0504130}}].

\bibitem{Bacchetta:2006tn}
A.~Bacchetta, M.~Diehl, K.~Goeke, A.~Metz, P.~J. Mulders, and M.~Schlegel, {\it
  {Semi-inclusive deep inelastic scattering at small transverse momentum}},
  {\em JHEP} {\bf 02} (2007) 093,
  [\href{http://arxiv.org/abs/hep-ph/0611265}{{\tt hep-ph/0611265}}].

\bibitem{Boer:2011xd}
D.~Boer, L.~Gamberg, B.~Musch, and A.~Prokudin, {\it {Bessel-Weighted
  Asymmetries in Semi Inclusive Deep Inelastic Scattering}},  {\em JHEP} {\bf
  10} (2011) 021, [\href{http://arxiv.org/abs/1107.5294}{{\tt
  arXiv:1107.5294}}].

\bibitem{Scimemi:2018mmi}
I.~Scimemi and A.~Vladimirov, {\it {Matching of transverse momentum dependent
  distributions at twist-3}},  {\em Eur. Phys. J. C} {\bf 78} (2018), no.~10
  802, [\href{http://arxiv.org/abs/1804.08148}{{\tt arXiv:1804.08148}}].

\bibitem{Boer:1997nt}
D.~Boer and P.~J. Mulders, {\it {Time reversal odd distribution functions in
  leptoproduction}},  {\em Phys. Rev. D} {\bf 57} (1998) 5780--5786,
  [\href{http://arxiv.org/abs/hep-ph/9711485}{{\tt hep-ph/9711485}}].

\bibitem{Rein:2022odl}
F.~Rein, S.~Rodini, A.~Sch{\"a}fer, and A.~Vladimirov, {\it {Sivers,
  Boer-Mulders and worm-gear distributions at next-to-leading order}},  {\em
  JHEP} {\bf 01} (2023) 116, [\href{http://arxiv.org/abs/2209.00962}{{\tt
  arXiv:2209.00962}}].

\bibitem{Sivers:1989cc}
D.~W. Sivers, {\it {Single Spin Production Asymmetries from the Hard Scattering
  of Point-Like Constituents}},  {\em Phys. Rev. D} {\bf 41} (1990) 83.

\bibitem{Collins:2002kn}
J.~C. Collins, {\it {Leading twist single transverse-spin asymmetries:
  Drell-Yan and deep inelastic scattering}},  {\em Phys. Lett. B} {\bf 536}
  (2002) 43--48, [\href{http://arxiv.org/abs/hep-ph/0204004}{{\tt
  hep-ph/0204004}}].

\bibitem{Brodsky:2002rv}
S.~J. Brodsky, D.~S. Hwang, and I.~Schmidt, {\it {Initial state interactions
  and single spin asymmetries in Drell-Yan processes}},  {\em Nucl. Phys. B}
  {\bf 642} (2002) 344--356, [\href{http://arxiv.org/abs/hep-ph/0206259}{{\tt
  hep-ph/0206259}}].

\bibitem{Brodsky:2002cx}
S.~J. Brodsky, D.~S. Hwang, and I.~Schmidt, {\it {Final state interactions and
  single spin asymmetries in semiinclusive deep inelastic scattering}},  {\em
  Phys. Lett. B} {\bf 530} (2002) 99--107,
  [\href{http://arxiv.org/abs/hep-ph/0201296}{{\tt hep-ph/0201296}}].

\bibitem{Brodsky:2013oya}
S.~J. Brodsky, D.~S. Hwang, Y.~V. Kovchegov, I.~Schmidt, and M.~D. Sievert,
  {\it {Single-Spin Asymmetries in Semi-inclusive Deep Inelastic Scattering and
  Drell-Yan Processes}},  {\em Phys. Rev. D} {\bf 88} (2013), no.~1 014032,
  [\href{http://arxiv.org/abs/1304.5237}{{\tt arXiv:1304.5237}}].

\bibitem{Belitsky:2002sm}
A.~V. Belitsky, X.~Ji, and F.~Yuan, {\it {Final state interactions and gauge
  invariant parton distributions}},  {\em Nucl. Phys. B} {\bf 656} (2003)
  165--198, [\href{http://arxiv.org/abs/hep-ph/0208038}{{\tt hep-ph/0208038}}].

\bibitem{Luo:2019hmp}
M.-X. Luo, X.~Wang, X.~Xu, L.~L. Yang, T.-Z. Yang, and H.~X. Zhu, {\it
  {Transverse Parton Distribution and Fragmentation Functions at NNLO: the
  Quark Case}},  {\em JHEP} {\bf 10} (2019) 083,
  [\href{http://arxiv.org/abs/1908.03831}{{\tt arXiv:1908.03831}}].

\bibitem{Metz:2016swz}
A.~Metz and A.~Vossen, {\it {Parton Fragmentation Functions}},  {\em Prog.
  Part. Nucl. Phys.} {\bf 91} (2016) 136--202,
  [\href{http://arxiv.org/abs/1607.02521}{{\tt arXiv:1607.02521}}].

\bibitem{Li:2016ctv}
Y.~Li and H.~X. Zhu, {\it {Bootstrapping Rapidity Anomalous Dimensions for
  Transverse-Momentum Resummation}},  {\em Phys. Rev. Lett.} {\bf 118} (2017),
  no.~2 022004, [\href{http://arxiv.org/abs/1604.01404}{{\tt
  arXiv:1604.01404}}].

\bibitem{Vogelsang:1997ak}
W.~Vogelsang, {\it {Next-to-leading order evolution of transversity
  distributions and Soffer's inequality}},  {\em Phys. Rev. D} {\bf 57} (1998)
  1886--1894, [\href{http://arxiv.org/abs/hep-ph/9706511}{{\tt
  hep-ph/9706511}}].

\bibitem{Mikhailov:2008my}
S.~V. Mikhailov and A.~A. Vladimirov, {\it {ERBL and DGLAP kernels for
  transversity distributions. Two-loop calculations in covariant gauge}},  {\em
  Phys. Lett. B} {\bf 671} (2009) 111--118,
  [\href{http://arxiv.org/abs/0810.1647}{{\tt arXiv:0810.1647}}].

\bibitem{Artru:1989zv}
X.~Artru and M.~Mekhfi, {\it {Transversely Polarized Parton Densities, their
  Evolution and their Measurement}},  {\em Z. Phys. C} {\bf 45} (1990) 669.

\bibitem{Barone:1997fh}
V.~Barone, {\it {On the QCD evolution of the transversity distribution}},  {\em
  Phys. Lett. B} {\bf 409} (1997) 499--502,
  [\href{http://arxiv.org/abs/hep-ph/9703343}{{\tt hep-ph/9703343}}].

\bibitem{Chiu:2011qc}
J.-y. Chiu, A.~Jain, D.~Neill, and I.~Z. Rothstein, {\it {The Rapidity
  Renormalization Group}},  {\em Phys. Rev. Lett.} {\bf 108} (2012) 151601,
  [\href{http://arxiv.org/abs/1104.0881}{{\tt arXiv:1104.0881}}].

\bibitem{Chiu:2012ir}
J.-Y. Chiu, A.~Jain, D.~Neill, and I.~Z. Rothstein, {\it {A Formalism for the
  Systematic Treatment of Rapidity Logarithms in Quantum Field Theory}},  {\em
  JHEP} {\bf 05} (2012) 084, [\href{http://arxiv.org/abs/1202.0814}{{\tt
  arXiv:1202.0814}}].

\bibitem{Blumlein:2009rg}
J.~Blumlein, S.~Klein, and B.~Todtli, {\it {O(alpha(s)**2) and O(alpha(s)**3)
  Heavy Flavor Contributions to Transversity at Q**2
  {\ensuremath{>}}{\ensuremath{>}}m**2}},  {\em Phys. Rev. D} {\bf 80} (2009)
  094010, [\href{http://arxiv.org/abs/0909.1547}{{\tt arXiv:0909.1547}}].

\bibitem{Ablinger:2014vwa}
J.~Ablinger, A.~Behring, J.~Bl{\"u}mlein, A.~De~Freitas, A.~Hasselhuhn, A.~von
  Manteuffel, M.~Round, C.~Schneider, and F.~Wi{\ss}brock, {\it {The 3-Loop
  Non-Singlet Heavy Flavor Contributions and Anomalous Dimensions for the
  Structure Function $F_2(x,Q^2)$ and Transversity}},  {\em Nucl. Phys. B} {\bf
  886} (2014) 733--823, [\href{http://arxiv.org/abs/1406.4654}{{\tt
  arXiv:1406.4654}}].

\bibitem{Luo:2019szz}
M.-x. Luo, T.-Z. Yang, H.~X. Zhu, and Y.~J. Zhu, {\it {Quark Transverse Parton
  Distribution at the Next-to-Next-to-Next-to-Leading Order}},  {\em Phys. Rev.
  Lett.} {\bf 124} (2020), no.~9 092001,
  [\href{http://arxiv.org/abs/1912.05778}{{\tt arXiv:1912.05778}}].

\bibitem{Vladimirov:2016dll}
A.~A. Vladimirov, {\it {Correspondence between Soft and Rapidity Anomalous
  Dimensions}},  {\em Phys. Rev. Lett.} {\bf 118} (2017), no.~6 062001,
  [\href{http://arxiv.org/abs/1610.05791}{{\tt arXiv:1610.05791}}].

\bibitem{Echevarria:2016scs}
M.~G. Echevarria, I.~Scimemi, and A.~Vladimirov, {\it {Unpolarized Transverse
  Momentum Dependent Parton Distribution and Fragmentation Functions at
  next-to-next-to-leading order}},  {\em JHEP} {\bf 09} (2016) 004,
  [\href{http://arxiv.org/abs/1604.07869}{{\tt arXiv:1604.07869}}].

\bibitem{Lustermans:2016nvk}
G.~Lustermans, W.~J. Waalewijn, and L.~Zeune, {\it {Joint transverse momentum
  and threshold resummation beyond NLL}},  {\em Phys. Lett. B} {\bf 762} (2016)
  447--454, [\href{http://arxiv.org/abs/1605.02740}{{\tt arXiv:1605.02740}}].

\bibitem{Billis:2019vxg}
G.~Billis, M.~A. Ebert, J.~K.~L. Michel, and F.~J. Tackmann, {\it {A Toolbox
  for $q_T$ and $0$-Jettiness Subtractions at N$^3$LO}},
  \href{http://arxiv.org/abs/1909.00811}{{\tt arXiv:1909.00811}}.

\bibitem{Buffing:2017mqm}
M.~G.~A. Buffing, M.~Diehl, and T.~Kasemets, {\it {Transverse momentum in
  double parton scattering: factorisation, evolution and matching}},  {\em
  JHEP} {\bf 01} (2018) 044, [\href{http://arxiv.org/abs/1708.03528}{{\tt
  arXiv:1708.03528}}].

\bibitem{Bacchetta:2013pqa}
A.~Bacchetta and A.~Prokudin, {\it {Evolution of the helicity and transversity
  Transverse-Momentum-Dependent parton distributions}},  {\em Nucl. Phys. B}
  {\bf 875} (2013) 536--551, [\href{http://arxiv.org/abs/1303.2129}{{\tt
  arXiv:1303.2129}}].

\bibitem{Moch:2017uml}
S.~Moch, B.~Ruijl, T.~Ueda, J.~A.~M. Vermaseren, and A.~Vogt, {\it {Four-Loop
  Non-Singlet Splitting Functions in the Planar Limit and Beyond}},  {\em JHEP}
  {\bf 10} (2017) 041, [\href{http://arxiv.org/abs/1707.08315}{{\tt
  arXiv:1707.08315}}].

\bibitem{Maitre:2005uu}
D.~Maitre, {\it {HPL, a mathematica implementation of the harmonic
  polylogarithms}},  {\em Comput. Phys. Commun.} {\bf 174} (2006) 222--240,
  [\href{http://arxiv.org/abs/hep-ph/0507152}{{\tt hep-ph/0507152}}].

\bibitem{Baikov:2016tgj}
P.~A. Baikov, K.~G. Chetyrkin, and J.~H. K\"uhn, {\it {Five-Loop Running of the
  QCD coupling constant}},  {\em Phys. Rev. Lett.} {\bf 118} (2017), no.~8
  082002, [\href{http://arxiv.org/abs/1606.08659}{{\tt arXiv:1606.08659}}].

\bibitem{Moch:2004pa}
S.~Moch, J.~A.~M. Vermaseren, and A.~Vogt, {\it {The Three loop splitting
  functions in QCD: The Nonsinglet case}},  {\em Nucl. Phys. B} {\bf 688}
  (2004) 101--134, [\href{http://arxiv.org/abs/hep-ph/0403192}{{\tt
  hep-ph/0403192}}].

\bibitem{Li:2014afw}
Y.~Li, A.~von Manteuffel, R.~M. Schabinger, and H.~X. Zhu, {\it {Soft-virtual
  corrections to Higgs production at N$^3$LO}},  {\em Phys. Rev. D} {\bf 91}
  (2015) 036008, [\href{http://arxiv.org/abs/1412.2771}{{\tt
  arXiv:1412.2771}}].

\bibitem{Moch:2005tm}
S.~Moch, J.~A.~M. Vermaseren, and A.~Vogt, {\it {Three-loop results for quark
  and gluon form-factors}},  {\em Phys. Lett.} {\bf B625} (2005) 245--252,
  [\href{http://arxiv.org/abs/hep-ph/0508055}{{\tt hep-ph/0508055}}].

\bibitem{Gehrmann:2010ue}
T.~Gehrmann, E.~W.~N. Glover, T.~Huber, N.~Ikizlerli, and C.~Studerus, {\it
  {Calculation of the quark and gluon form factors to three loops in QCD}},
  {\em JHEP} {\bf 06} (2010) 094, [\href{http://arxiv.org/abs/1004.3653}{{\tt
  arXiv:1004.3653}}].

\bibitem{Becher:2009qa}
T.~Becher and M.~Neubert, {\it {On the Structure of Infrared Singularities of
  Gauge-Theory Amplitudes}},  {\em JHEP} {\bf 06} (2009) 081,
  [\href{http://arxiv.org/abs/0903.1126}{{\tt arXiv:0903.1126}}]. [Erratum:
  JHEP11,024(2013)].

\bibitem{Manohar:2006nz}
A.~V. Manohar and I.~W. Stewart, {\it {The Zero-Bin and Mode Factorization in
  Quantum Field Theory}},  {\em Phys. Rev.} {\bf D76} (2007) 074002,
  [\href{http://arxiv.org/abs/hep-ph/0605001}{{\tt hep-ph/0605001}}].

\end{thebibliography}\endgroup

\end{document}